\documentclass[11pt, epsfig]{article}
\usepackage{epsfig, amsmath, amssymb, amsthm, times}
\usepackage{graphicx}
\usepackage{color}
\usepackage{bm}% bold math

\newcommand{\R}{\mathbb{R}}
\newcommand{\T}{\mathbb{T}}
\newcommand{\C}{\mathbb{C}}
\newcommand{\Z}{\mathbb{Z}}

\newcommand{\bW}{\boldsymbol{W}}
\newcommand{\bT}{\boldsymbol{T}}

\newcommand{\cC}{\mathcal{C}}

\newcommand{\cH}{\mathcal{H}}

\newcommand{\bbH}{\mathbb{H}}

\def\lbl{\label}
\def\be{\begin{equation}}
\def\ee{\end{equation}}
\def\p{\partial}

\newcommand{\iu}{{i\mkern1mu}}

\def\1{\mathbf{1}}

\renewcommand{\Re}{\operatorname{Re}}
\renewcommand{\Im}{\operatorname{Im}}

\title{Chimeras unfolded}
\author{Georgi S. Medvedev\thanks{Department of Mathematics, 
		Drexel University, 3141 Chestnut Street, Philadelphia, PA 19104,
		{\tt medvedev@drexel.edu}} \and Matthew S. Mizuhara\thanks{Department of 
		Mathematics and Statistics,
		The College of New Jersey,
		{\tt  mizuharm@tcnj.edu}}}
\begin{document}
\maketitle
\begin{abstract}
  The instability of mixing in the Kuramoto model of coupled phase oscillators is the
  key to understanding a range of spatiotemporal patterns, which feature prominently
  in collective dynamics of systems ranging from neuronal networks, to coupled lasers,
  to power grids.

  In this paper, we describe a codimension--$2$ bifurcation of mixing
  whose unfolding, in addition to the classical scenario of the onset
  of synchronization, also  explains the formation of clusters and chimeras.
  We use a combination of linear stability analysis and Penrose diagrams
  to identify and analyze a variety of spatiotemporal patterns including
  stationary and traveling coherent clusters and twisted states, as well as their
  combinations with regions of incoherent behavior called chimera states.
  The linear stability analysis is used to
  estimate of the velocity distribution within these structures. Penrose diagrams,
  on the other hand, predict accurately the basins of
  their existence. Furthermore, we show that network topology can endow chimera states
  with nontrivial spatial organization. In particular, we present twisted chimera states, whose
  coherent regions are organized as stationary or traveling twisted states.
   The analytical results are illustrated with numerical bifurcation diagrams computed for the Kuramoto model
  with uni-, bi-, and tri- modal   frequency distributions and all-to-all and nonlocal nearest-neighbor
  connectivity.
\end{abstract}

\section{Introduction}
\setcounter{equation}{0}
The Kuramoto model (KM) on a graph sequence $(\Gamma^n)$ describes collective dynamics in
coupled networks. It is given by the following system of ordinary
differential equations
\begin{equation}\label{KM}
  \dot \theta_i=\omega_i +2Kn^{-1} \sum_{j=1}^n
  a^n_{ij}\sin(\theta_j-\theta_i+\alpha), \quad  i\in [n]:=\{1,2,\dots,n\}.
\end{equation}
Here,  $\theta_i:\R^+\to \T:=\R/2\pi\Z,$
stands for the phase of oscillator $i$;
$\omega_i$'s are independent random intrinsic frequencies drawn
from a distribution
with density $g(\omega),$  and $K$ is the strength of coupling. 
 $(a^n_{ij})$ is the
$n\times n$
symmetric (weighted) adjacency matrix of $\Gamma^n$, a graph
on $n$ nodes, which defines the connectivity of the network.
 The phase lag $\alpha$ controls the type of coupling and
 can play a role in pattern formation \cite{AbrStr06, Ome18}. It will not be used
 in this work and, thus, $\alpha$ is set to $0$.

Despite its analytical simplicity, the KM provides important insights into general
principles underlying network dynamics.
It is best known for revealing a universal scenario of  transition to synchronization,
identified in a variety of systems from neuronal networks, to coupled lasers,
to power grids \cite{Str00, Rod16}. More recently, the KM became the main framework for studying chimera
states, counterintuitive patterns combining regions of coherent and incoherent dynamics
\cite{KurBat02, AbrStr06, OmeMai08, Laing09, Ome13, PanAbr15, Ome18}.
For a long time, coherence and incoherence were viewed as distinct regimes in network
dynamics. Computational and experimental studies of chimeras clearly demonstrate that
coexistence of coherence and incoherence is ubiquitous in
diverse physical and biochemical systems \cite{ChiExp-1,ChiExp-2, ChiExp-3,PanAbr15, Ome18}.
Since the discovery of chimera states by Kuramoto and Battogtokh in 2002 \cite{KurBat02},
there has been a continuous stream of papers suggesting different dynamical mechanisms
for their generation. Many such studies rely heavily on numerical
simulations. The most complete analytical information about chimera states was
derived using the Ott-Antonsen Ansatz \cite{OttAnt08, AbrStr08, Laing09, Ome13, Ome18},
which exploits the symmetries of the KM. When applicable the
Ott-Antonsen Ansatz provides a powerful tool
for studying chimera states. However, not all chimera states lie in the Ott-Antonsen manifold
(see \cite{Ome18} for a discussion of the benefits and limitations of the Ott-Antonsen Ansatz).

In the present paper, we describe a bifurcation scenario, which connects mixing to clusters
to chimeras. At the heart of this scenario lies a codimension-2 bifurcation of mixing, whose
unfolding contains clusters and chimeras. We use the linear stability analysis of mixing \cite{ChiMed19a}
and Penrose diagrams \cite{Penrose60} to locate different bifurcations and to describe  statistical
properties of the patterns emerging from them. We relate the pitchfork (PF) and Andronov-Hopf
(AH) bifurcations to the appearance of synchronized stationary and traveling clusters.
The eigenfunctions of the linearized operator
corresponding to bifurcating eigenvalues capture the velocity distributions within partially locked
states (PLS) and chimera states emerging when mixing loses stability.
Furthermore, Penrose diagrams provide a crisp
picture of the bifurcation scenarios in the KM and accurately  predict the domain
of existence of chimera states. 

After a brief review of the linear stability analysis of mixing following \cite{ChiMed19a, CMM20} in
Section~\ref{sec.linear}, we turn to the analysis of bifurcations in the KM with uni-, bi- and tri- modal
frequency distribution in Section~\ref{sec.bif}.  We start with the unimodal distribution to explain how to use Penrose
diagrams to locate the bifurcation in the KM \cite{Die16}. Then we apply this method to study
bifurcations in the KM with
bimodal friquency distributions. Here, we identify the codimension-2 PF-AH
bifurcation, which is responsible for the emergence of clusters in the KM model. By breaking
symmetry of the bimodal frequency distribution, we locate chimeras and  clusters bifurcating from mixing.
To illustrate
bifurcations in the KM with multimodal distributions more fully we also discuss bifurcations
in the KM with trimodal frequency distributions.
Here, as in the bimodal case, we identify a master bifurcation of mixing, whose unfolding
contains bifurcation scenarios connecting mixing to chimeras and three-cluster states.
In Section~\ref{sec.connect}, we address the effects
of the network connectivity on spatial organization of patterns emerging from the bifurcations of mixing.
To this end, we show that nonlocal nearest-neighbor connectivity transforms the clusters of synchronized
behavior into clusters of twisted states. In particular, we demonstrate various patterns involving stationary
and traveling twisted states, as well as twisted chimera states. We conclude with a brief discussion of the
results of this work in Section~\ref{sec.discuss}.

\section{Stability of mixing}\label{sec.linear}
\setcounter{equation}{0}
A starting point in virtually any approach to the analysis of chimera states is
 the thermodynamic limit as the size of the system $n$ tends to $\infty$.
Clearly, to expect a common limiting behavior of solutions of the discrete problems
\eqref{KM}, the corresponding graphs $(\Gamma^n)$ need to have a well defined asymptotic
behavior as $n\to\infty$ as well. To this end, we assume that $(\Gamma^n)$ is a
convergent sequence of dense graphs, whose  limit is given by graphon
$W:[0,1]^2\rightarrow [0,1]$. Graphons are symmetric measurable functions representing
graphs and graph limits \cite{LovGraphLim12}. More details on using graphons
in dynamical models can be found in \cite{ChiMed19a, Med19}.

For the model at hand, the thermodynamic limit is given by the following Vlasov equation 
\begin{equation}\label{MF}
\partial_t f(t,\theta,\omega,x) +
\partial_\theta \left\{ V(t,\theta,\omega,x) f(t,\theta,\omega,x) \right\}=0,
\end{equation}
where $f(t,\theta,\omega,x)d\theta d\omega$ is the probability
that the state of the oscillator at point $x\in I:=[0,1]$ at time $t\in \R^+$ is in
$(\theta, \theta+d\theta)\times (\omega, \omega+d\omega)$.
The velocity field is derived from the right--hand side of \eqref{KM}:
\begin{equation}\lbl{def-V}
  V(t,\theta,\omega,x) =\omega- K\iu
                        \left( \kappa(t,x) e^{-\iu\theta} -\overline{\kappa(t,x)
                            e^{-\iu\theta}}\right),
                      \end{equation}
                      where
                      \begin{eqnarray}\lbl{corder}
  \kappa(t,x)  &=&\int_\T e^{\iu\theta} \int_\R
                \left(\bW f(t,\theta,\omega,\cdot) \right)(x) d\omega d\theta,  \\            
\lbl{def-W}
\left(\bW \phi \right)(x)  &=&\int_I W(x,y)\phi(y)dy.
\end{eqnarray}
The function $\kappa(t,x)$  is called a local order parameter. It
 provides a convenient measure of coherence in network dynamics near
 $x\in I$ at time $t$. 
 The self--adjoint operator $\bW$ is determined by $W$, which in turn
 reflects the asymptotic connectivity of the network.             
             A rigorous justification of the mean field limit \eqref{MF} in the context of the KM with all--to--all
             coupling was given in \cite{Lan05}. For the KM on convergent graph sequences, the use
             of the Vlasov equation as a mean field limit was further
justified in \cite{KVMed18, ChiMed19a}.
% For the KM on sparse graphs with unbounded degree, the results  in \cite{KVMed18, ChiMed19a}
% continue to hold when combined with \cite[Theorem~4.1]{Med19}.

Equation \eqref{MF} has a steady state solution
\begin{equation}\label{steady}
  f_{mix}={g(\omega)\over 2\pi},
\end{equation}
which is called mixing.
It corresponds to the uniform distribution of the phases over $\T$.
Stability of mixing for the KM on graphs was analyzed in \cite{ChiMed19a}.
For completeness, we outline the main steps of the linear stability analysis
below.

It is convenient to study stability of $f_{mix}$ in the Fourier space. To this end, we
introduce
\begin{equation}\lbl{FT}
u_l(t,\omega, x)= \int_\T  e^{\iu l\theta} f(t,\theta, \omega, x) d\theta,
                 \quad l\in \Z.
               \end{equation}
               Note that
               \begin{equation}\label{marginal}
                  u_0(t,\omega,x) =\int_\R  f(t,\theta, \omega, x) d\theta =g(\omega).
                \end{equation}
                By applying the Fourier transform to \eqref{MF} and keeping only the linear terms, we have
\begin{eqnarray}\lbl{u1}
  \p_t u_1(t,\omega,\cdot) &=& \iu \omega u_1(t,\omega, \cdot) +
                               K g(\omega) \bW \left[ \int_R u_1(t,\omega, \cdot) d\omega\right],\\
\lbl{ul}
  \p_t u_l(t,\omega, \cdot) &=& \iu l  \omega u_l(t,\omega, \cdot), \quad l\ge 1.
\end{eqnarray}
It was sufficient to restrict to $l\ge 1$ in \eqref{ul}, because $f$ is real and, thus, 
$u_{-l} =\bar u_l$.

Further, the steady state $f_{mix}$ is mapped to $u_{mix}=(g(\omega), 0,0,\dots)$ in the Fourier space.
Equations in \eqref{ul} describe pure transport. Thus, the stability of $u_{mix}$ is decided by \eqref{u1},
which we rewrite as
\begin{equation}\label{Linear}
  \partial_t v =\bT[v],
\end{equation}
where
\begin{align}
  \lbl{def-T}
  \bT[\upsilon] (\omega, \cdot)  =\iu \omega \upsilon + K g(\omega)
  \bW \left[ \int_R \upsilon (\xi, \cdot) d\xi\right], \quad v\in
  \cH=L^2\left( \R\times [0,1]\right).
\end{align}
As an operator on $\cH$, $\bT$ has continuous spectrum on $\iu\R$ (cf. \cite{ChiMed19a}). To locate the
eigenvalues of $\bT$, we consider the following spectral problem
\begin{equation}\label{spectral}
  \bT[\upsilon]=\lambda\upsilon,\quad \upsilon\in\cH.
\end{equation}
Using \eqref{def-T}, we rewrite \eqref{spectral} as follows
\begin{equation}\label{re-spectral}
  K\frac{g(\omega)}{\lambda-\iu \omega}
  \bW \left[ \int_\R \upsilon (\xi,\cdot) d\xi\right]=\upsilon (\omega,\cdot).
\end{equation}
By integrating both parts of \eqref{re-spectral} over $\R$, we arrive at
\begin{equation}\label{int-spectral}
  KG(\lambda) \bW[w]=w, 
\end{equation}
where $w=\int_R \upsilon (\xi,\cdot) d\xi\in L^2([0,1])$ and
\begin{equation}\label{def-G}
  G(\lambda)=\int_\R \frac{g(\omega) d\omega}{\lambda-\iu \omega}.
\end{equation}

As a  compact self--adjoint operator on $ L^2([0,1])$, $\bW$ has a
countable sequence of eigenvalues with a single accumulation point at $0$.
Let $\mu$ be an arbitrary fixed nonzero eigenvalue of $\bW$ and let
$w_\mu$ be a corresponding eigenfunction. From \eqref{int-spectral}, we find
the following equation for the eigenvalues of $\bT$:
\begin{equation}\label{key-eqn}
  G(\lambda)=\frac{1}{K\mu}.
\end{equation}
A root of \eqref{key-eqn} is an eigenvalue of $\bT$. The  corresponding
eigenfunction is then found from \eqref{re-spectral}
\begin{equation}
\label{v-lam}
           v_\lambda(\omega,x)=
           \Upsilon_\lambda(\omega) w_\mu(x), \quad \Upsilon_\lambda (\omega)=
           {g(\omega)\over \lambda-\iu\omega}.
\end{equation}

For $\lambda\notin \iu\R$, $\Upsilon_\lambda$ is a holomorphic function.
Since we are interested in bifurcations of mixing, 
we need to
resolve the meaning of $\Upsilon_\lambda$ for $\lambda\in \iu\R$. To this end, we impose
the following assumptions on the class of admissible probability density functions $g\in L^1([0,1])$.
Following \cite{Die16}, we assume that the Fourier transform of $g$, $\hat g\in C(\R)$ and
\begin{equation}\label{eat}
  \sup_{t\in \R} e^{at}\left| \hat g(t)\right| <\infty
\end{equation}
for some $a>0$.
Under these assumptions, $\Upsilon_{\iu y}$ can be viewed as a tempered distribution \cite{CMM20}.
Specifically, for any $\phi$ from the Schwartz
           class $\mathcal{S}(\R)$, by Sokhotski--Plemelj formula (cf.~\cite{Simon-Complex}), we have
           \begin{alignat*}{2}
           \langle  \Upsilon_{\iu y},\phi\rangle & =\lim_{\lambda \to \iu y+0}
           \int_{-\infty}^\infty {g(\omega)\phi(\omega)\over \lambda+\iu y-\iu\omega}d\omega\\
           &=\lim_{\lambda \to 0+}
           \int_{-\infty}^\infty {g(\omega+y)\phi(\omega+y)\over \lambda-\iu\omega}d\omega\\
&=\pi g(y)\phi(y)-\iu\operatorname{pv}
           \int_{-\infty}^\infty {g(\omega+y)\phi(\omega+y)\over \omega} d\omega.
           \end{alignat*}
           Thus, $\Upsilon_{\iu y} \in\mathcal{S}^\prime (\R)$ and 
          \be\lbl{viy-distribution}
    \Upsilon_{\iu y} =
           \pi g(y) \delta_{y}-\iu\mathcal{P}_{y}[g],
           \ee
           where $\delta_y$ stands for the Dirac's delta function centered at $y$ and
           $$
           \langle \mathcal{P}_y[g], \phi \rangle=\operatorname{pv}
             \int_{-\infty}^\infty {g(\omega+y)\phi(\omega+y)\over \omega}d\omega.
           $$
In particular,     
            \be\lbl{v0-distribution}
            \Upsilon_0=\pi g(0) \delta_0-\iu\mathcal{P}_0[g],
           \ee
           The eigenfunctions \eqref{v-lam}, \eqref{viy-distribution} corresponding to bifurcating
           eigenvalues will be used to explain spatiotemporal  patterns arising at the loss of stability
           of mixing and solutions bifurcating from mixing.
       
\section{Bifurcations: from mixing to chimeras}\label{sec.bif}
\setcounter{equation}{0}
In this section, we describe a sequence
of bifurcations from mixing to clusters to chimeras. We will show that all these structures
belong to the
unfolding of a codimension-$2$ bifurcation of mixing, the central object of our study.
In the following section, we will discuss 
the role of $W$ in shaping chimera states.
Until then, we restrict to $W\equiv 1$. This corresponds to all--to--all connectivity.
The bifurcation scenarios discussed below will also hold for the KM on any
graph sequence with a constant graph limit, e.g.,   Erd\H{o}s-R{\' enyi} or Paley graphs \cite{CMM18}.
For $W\equiv 1$, the only nonzero eigenvalue of $\bW$ is $\mu=1$. 
Thus, the equation for the eigenvalues of $\bT$ \eqref{key-eqn} takes the following form
\begin{equation}\label{GK}
  G(z)=K^{-1},
\end{equation}
where $G$ is defined in \eqref{def-G}.

A rigorous analysis of bifurcations in this model requires the generalized spectral theory
\cite{Chi15b}. This is related to the fact that as an operator on $L^2(\R\times [0,1])$,
$\bT$ has continuous spectrum filling the imaginary axis.
Thus, to be able to trace the eigenvalues
crossing the imaginary axis under the variation of $K$,
$\bT$ has to be given a more general interpretation. For the KM on graphs this was done in \cite{Chi15b}.
In the present paper, to avoid
the technicalities of the generalized spectral theory, we take the following approach.
We locate the eigenvalues of $\bT$ in $\bbH^+=\{z\in \C:\; \Re z>0\}$, where the corresponding
eigenfunctions are still in $L^2([\R\times [0,1])$. Then we trace these eigenvalues
until they hit the imaginary axis and identify the corresponding bifurcations.
As soon as the spectral parameter $\lambda$ hits the imaginary axis, the corresponding
eigenfunctions leave $L^2(\R\times [0,1])$. From this point, they are interpreted as tempered distributions
(cf.~\eqref{viy-distribution}).
We relate these eigenfunctions to the patterns emerging at bifurcations of mixing. In particular,
we show that a PF bifurcation results in a PLS with stationary coherent cluster, whereas an AH bifurcation
leads to the creation of two moving clusters.

\begin{figure}
  \centering
\textbf{a}\includegraphics[width=0.33\textwidth]{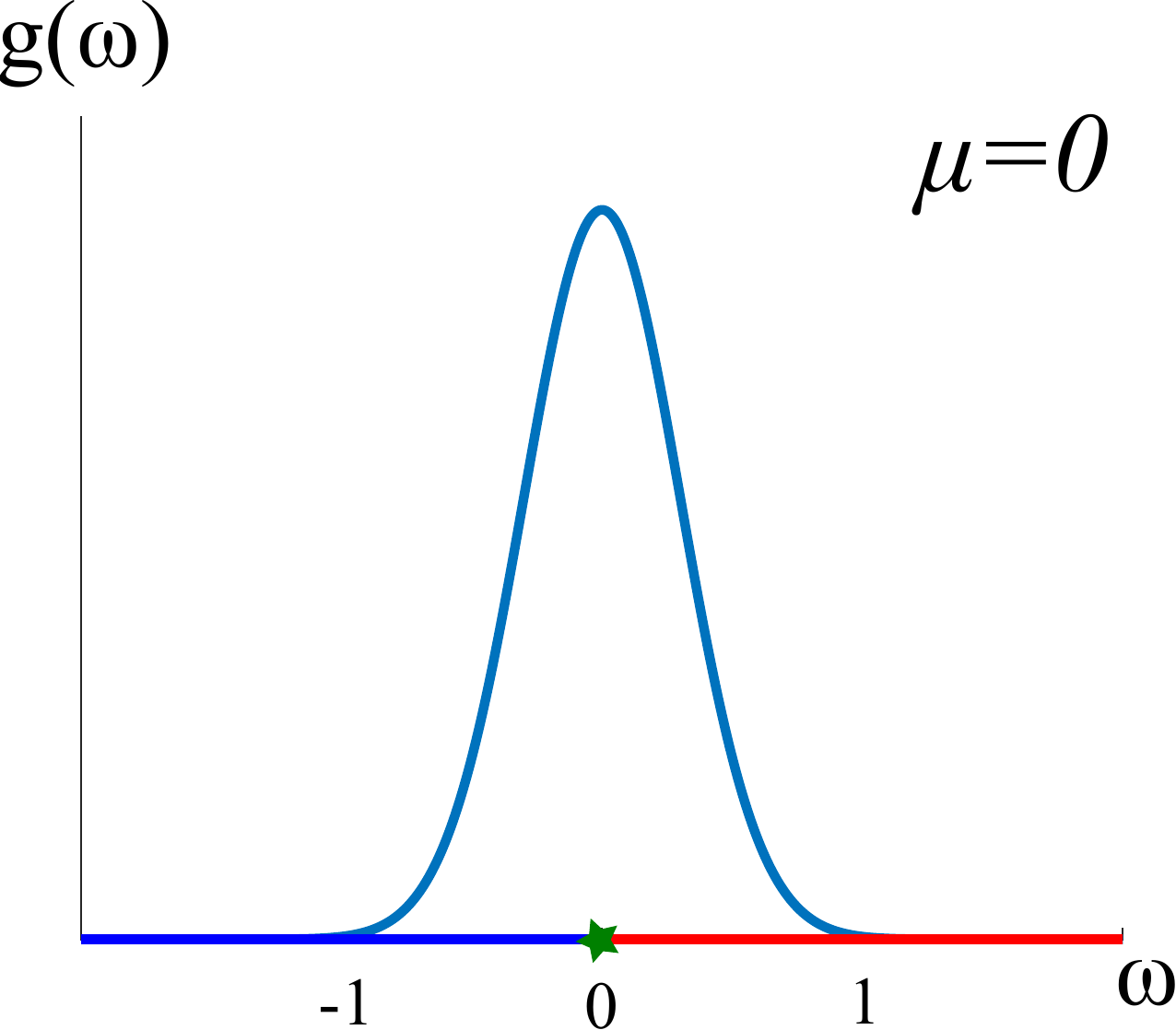}
\quad
\textbf{b}\includegraphics[width=0.33\textwidth]{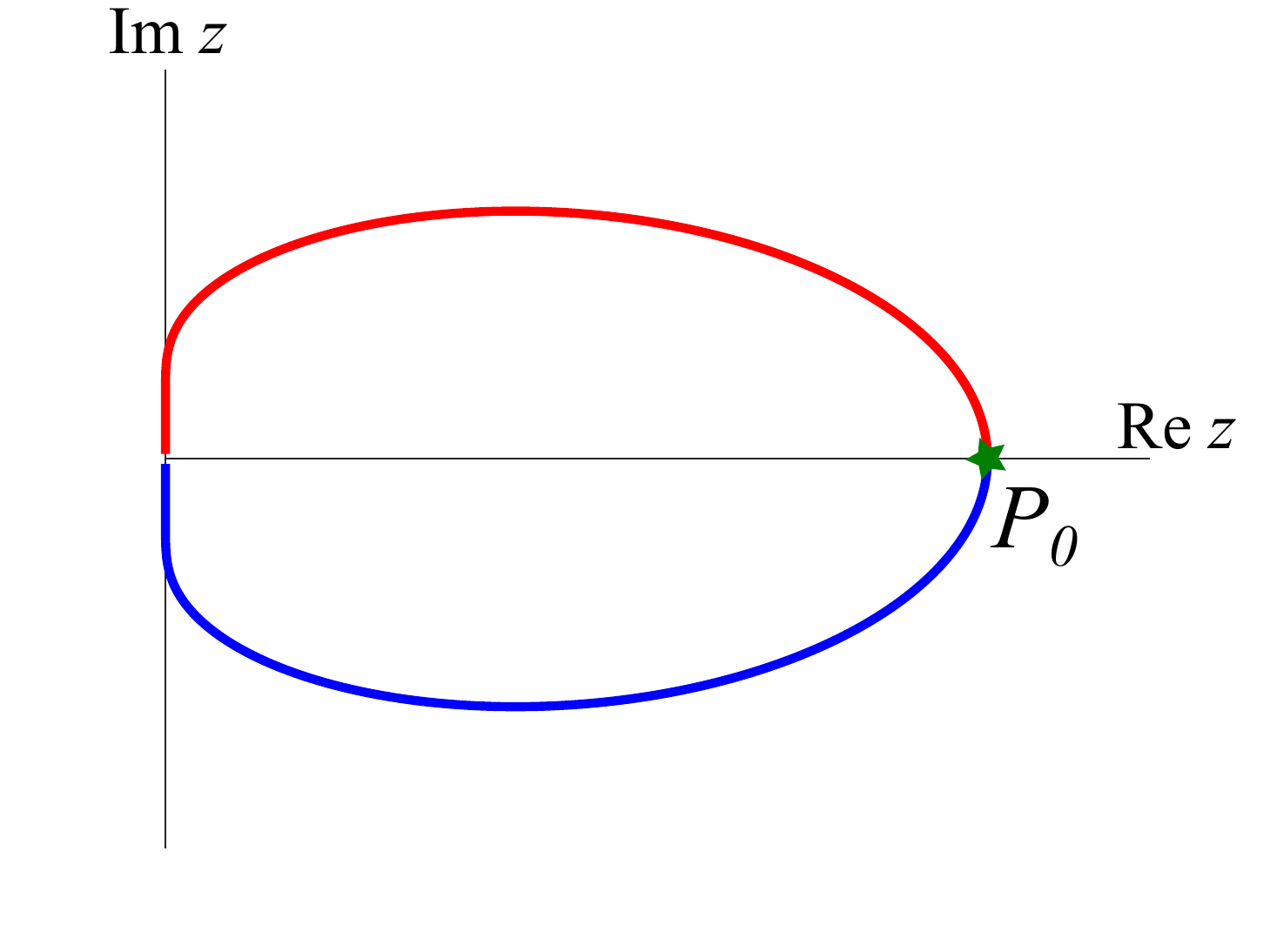}\\
\textbf{c}\includegraphics[width=0.33\textwidth]{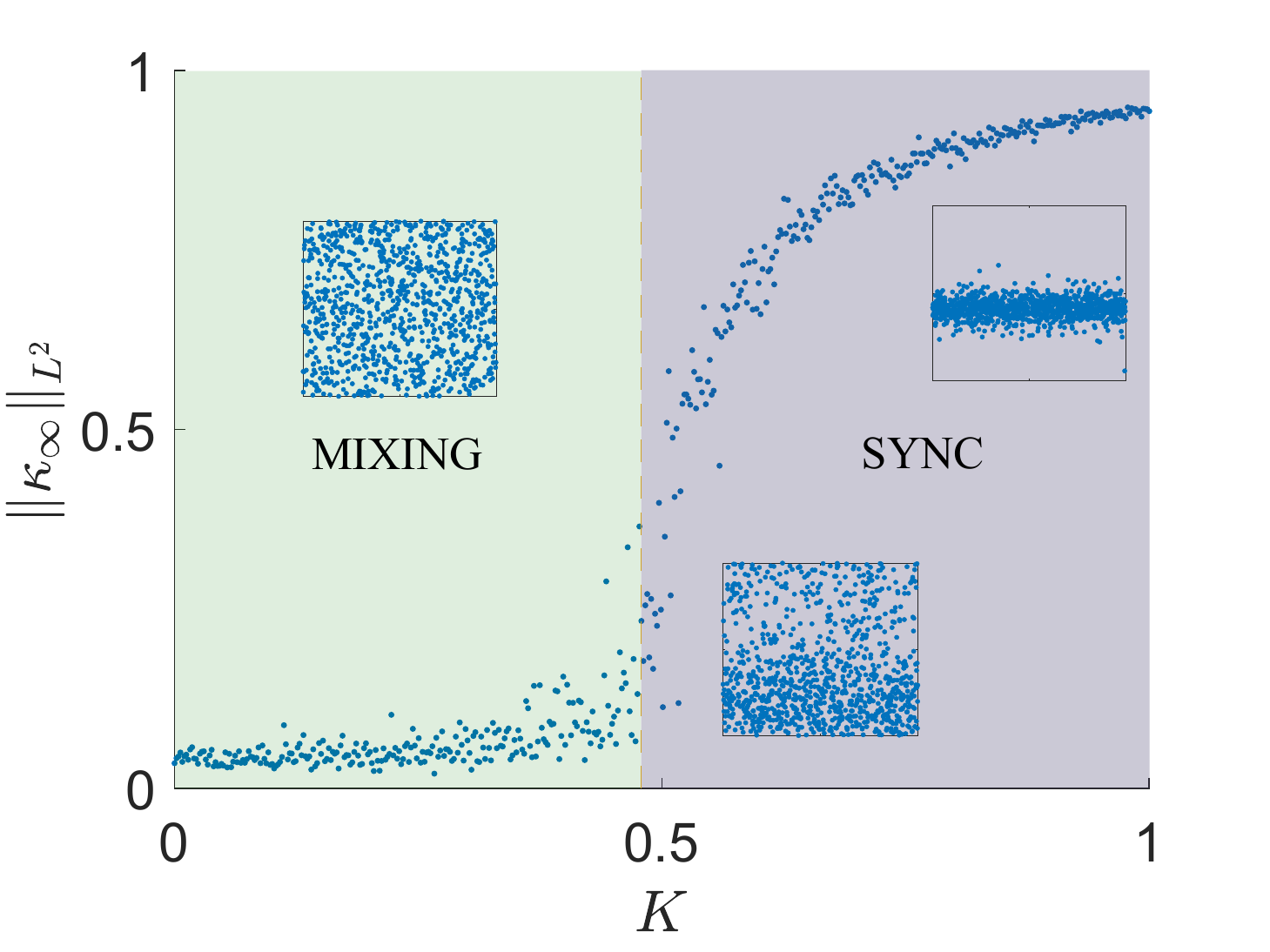}
\quad
\textbf{d}\includegraphics[width=0.33\textwidth]{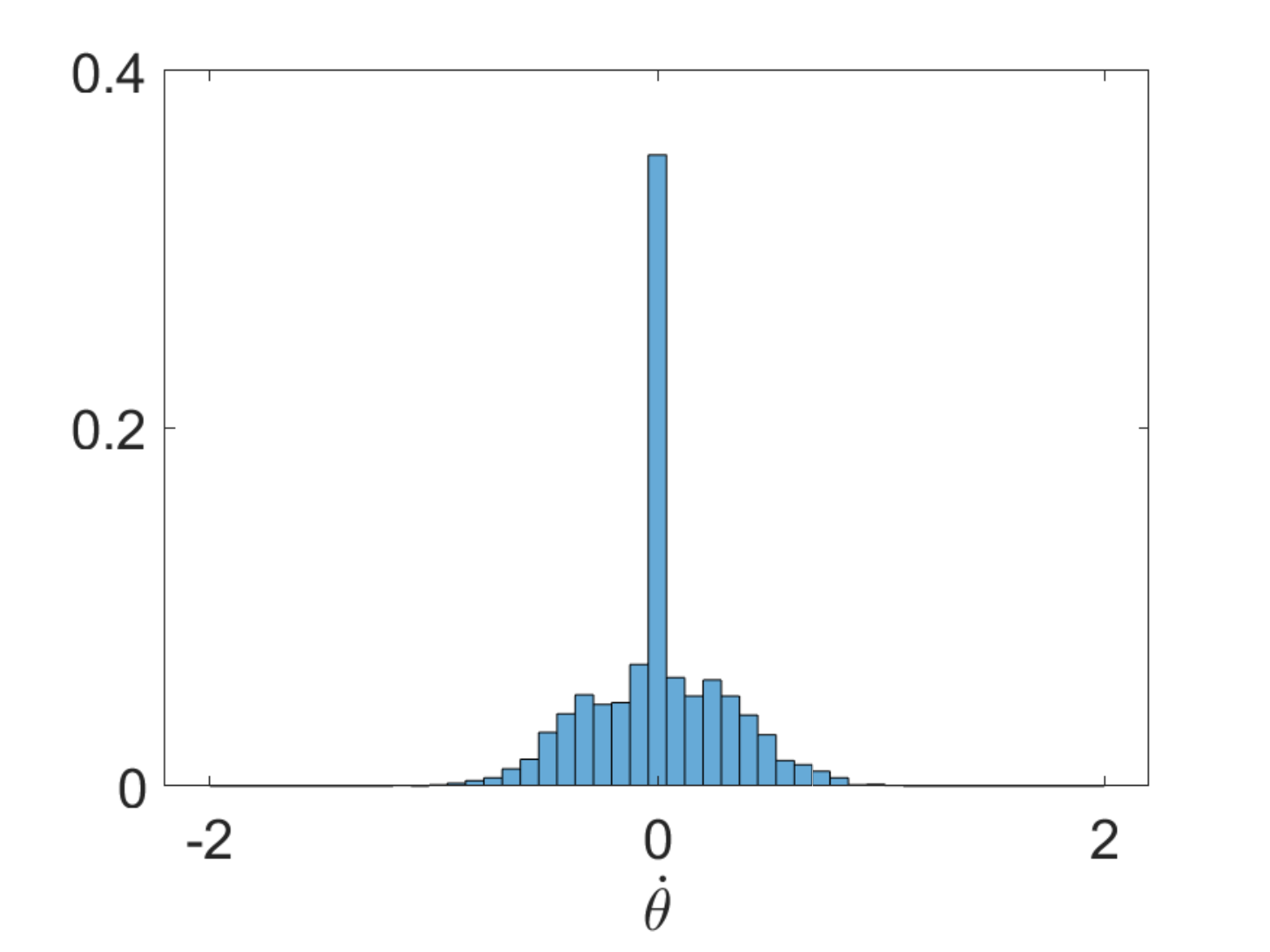}
\caption{\textbf{a}) A graph of an even unimodal probability density function $g$.
 \textbf{b}) The corresponding critical curve $\cC$ intersects postive
 real semiaxis at a unique point $P_0$. The preimage of $P_0$ under $G$ is indicated by
 the green star in (\textbf{a}). $P_0$ corresponds to the PF bifurcation of mixing resulting in a
PLS, which is then gradually transformed into
  synchronous state (see (\textbf{c})). \textbf{d}) The velocity distribution within the PLS
  near PF bifurcation is determined by the eigenfunction \eqref{vPF}.
  The normalized histogram of the velocities within the PLS.
}
\label{f.uni}
\end{figure}

To locate the roots of \eqref{GK} in $\bbH^+$ we employ the
method used by Penrose in \cite{Penrose60}. To this end, let $\cC$ denote an oriented curve
$G(\iu t),$ $t\in\R$. First, we establish certain  qualitative properties of $\cC$. To this end, we note that
$G$ is a holomorphic function on $\bbH^+$ (cf.~\eqref{def-G}). By the Paley-Wiener theorem, using
\eqref{eat}, $G$ can be extended analytically to the region $\Im z>-a$. Further, we use
the Sokhotski-Plemelj formula \cite{Simon-Complex} to obtain
\begin{equation}\label{SP}
  G(\iu t+0)=\pi g(t)+\iu \int_0^\infty \frac{g(t+s)-g(t-s)}{s} ds.
\end{equation}
This yields the following parametric equations for $\cC$:
\begin{equation}\label{parC}
  \begin{split}  x&=\pi g(t), \\
    y&=\int_0^\infty \frac{g(t+s)-g(t-s)}{s} ds,
    \end{split}
\end{equation}
for $t\in \R$. It follows from \eqref{parC} that $\cC$ lies in $\overline{\bbH}^+$ and
$\cC$ asymptotes onto the origin. 
Thus, $\cC$ is a bounded closed curve  in $\overline{\bbH}^+$ (see Fig.~\ref{f.uni}\textbf{b}).

\subsection{Unimodal $g$}
\begin{figure}
  \centering
\begin{minipage}[b]{.98\textwidth}
\textbf{a}\includegraphics[width=0.3\textwidth]{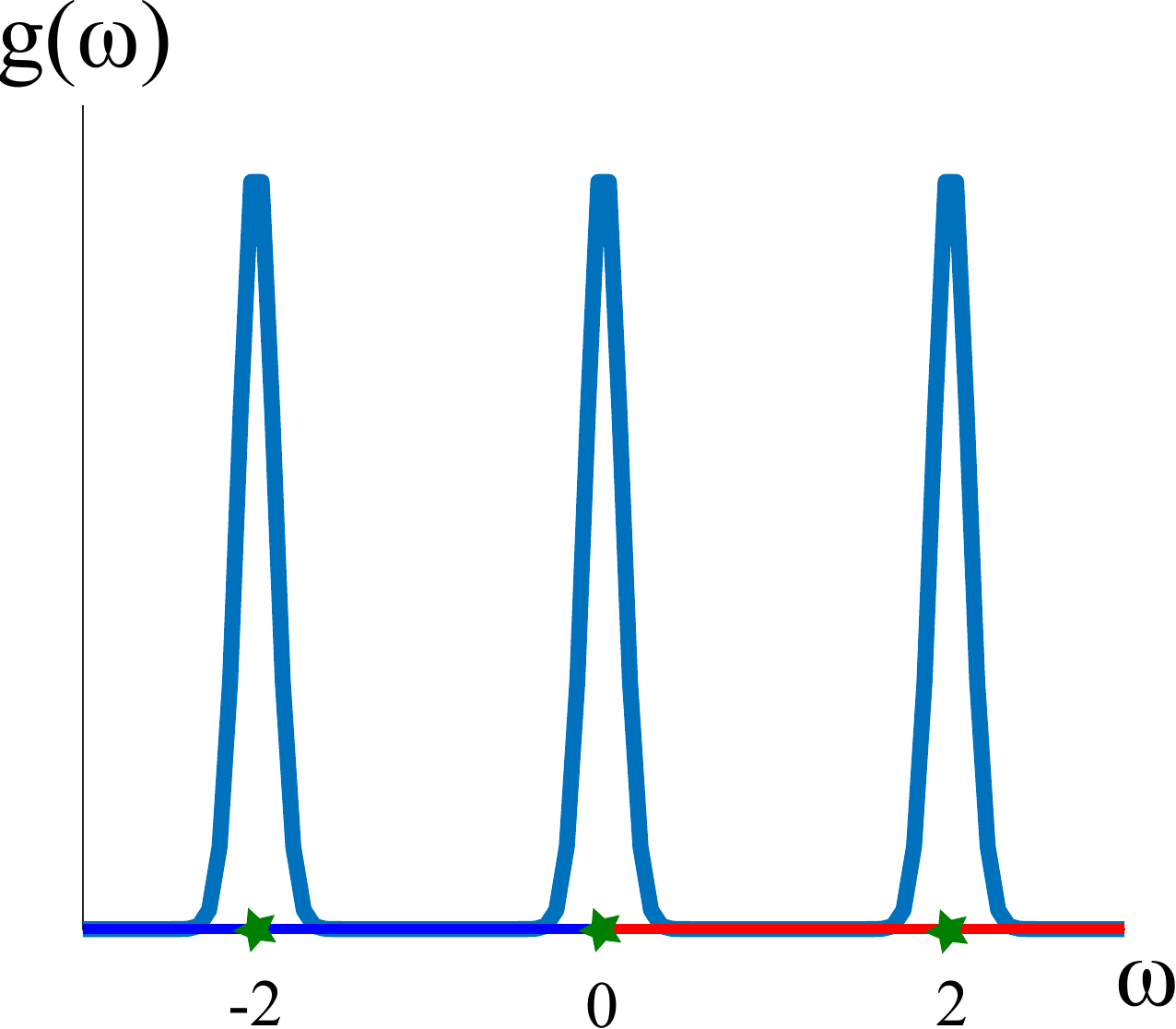}
\hfill
\textbf{b}\includegraphics[width=0.3\textwidth]{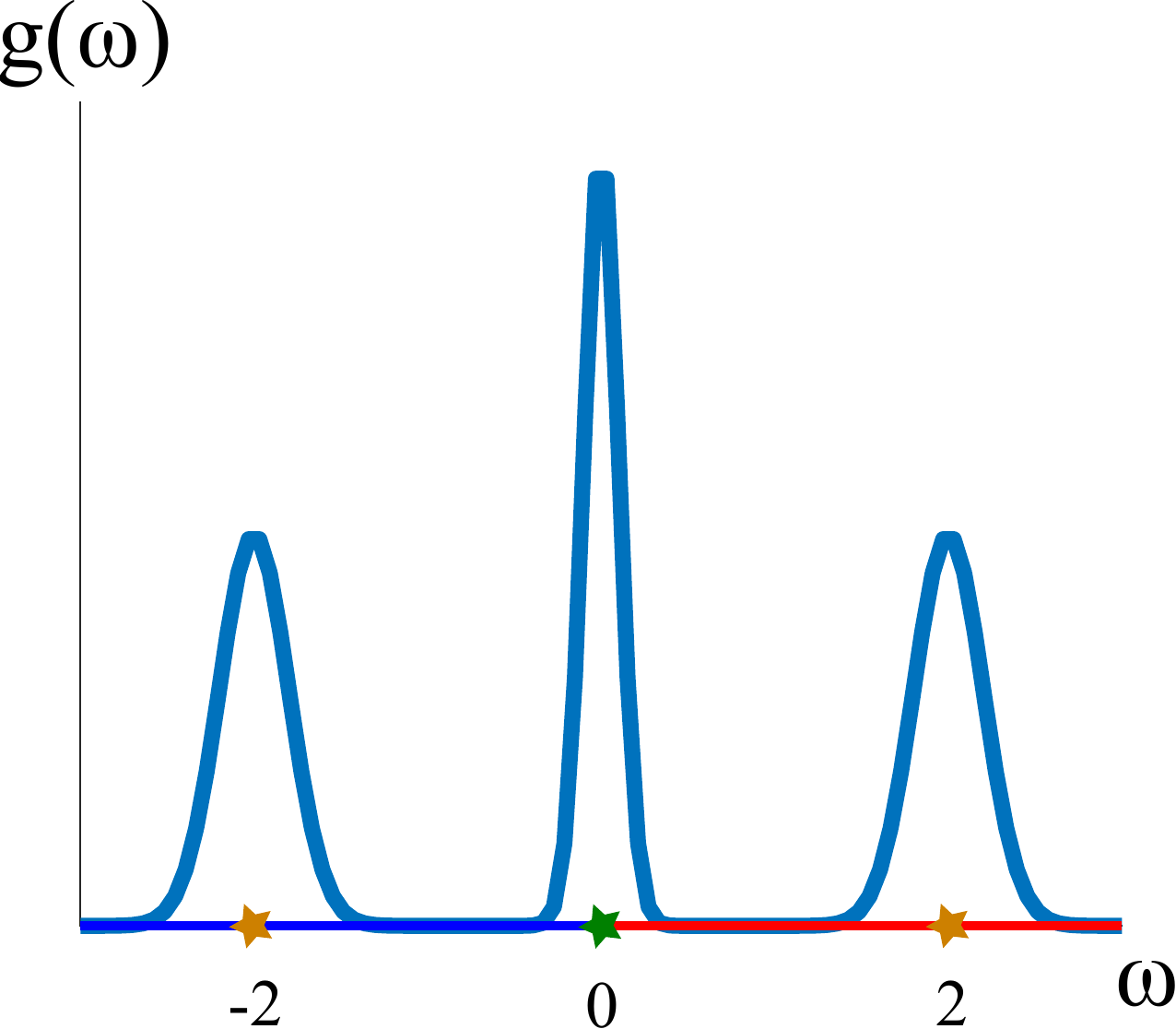}
\hfill
\textbf{c} \includegraphics[width = .31\textwidth]{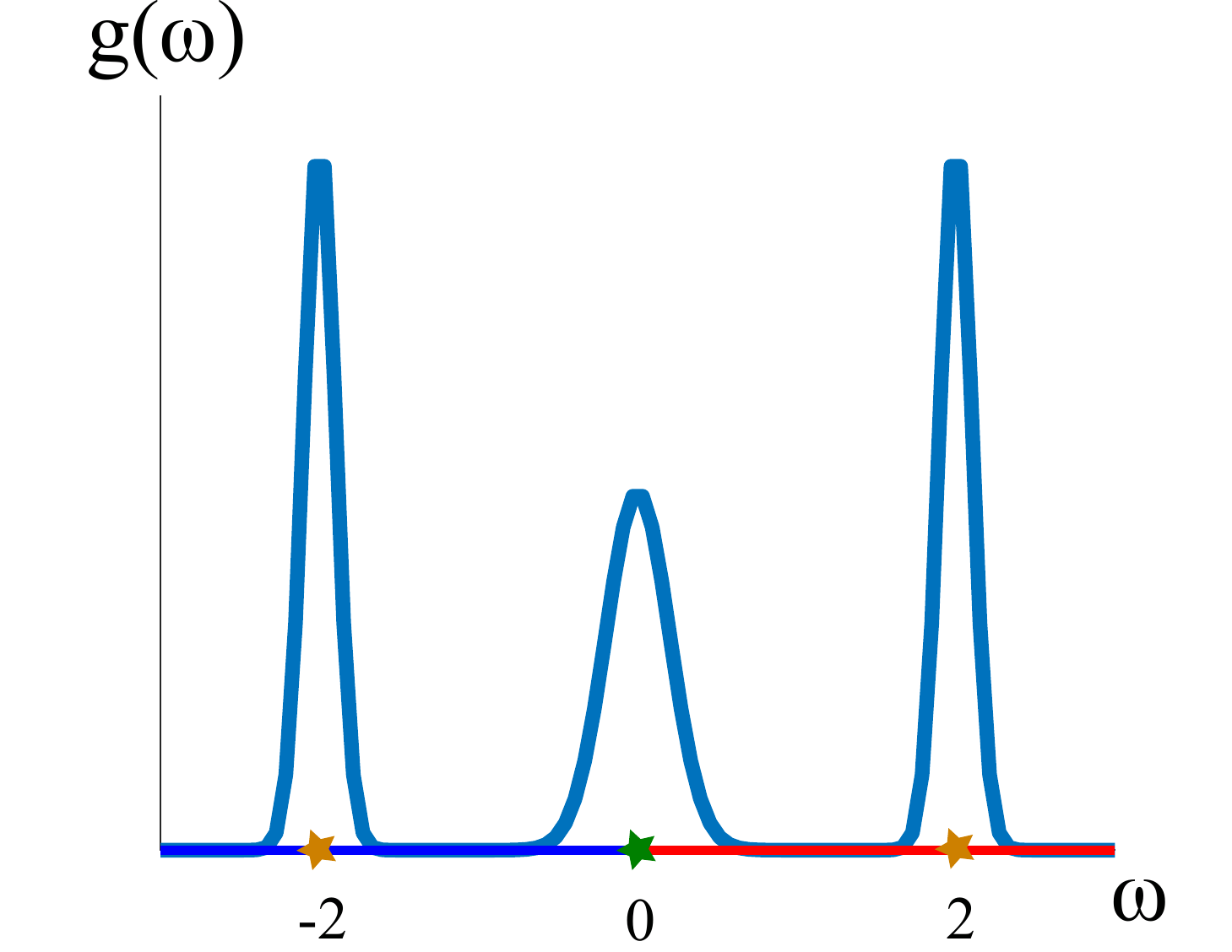} \\
\textbf{d}\includegraphics[width=0.3\textwidth]{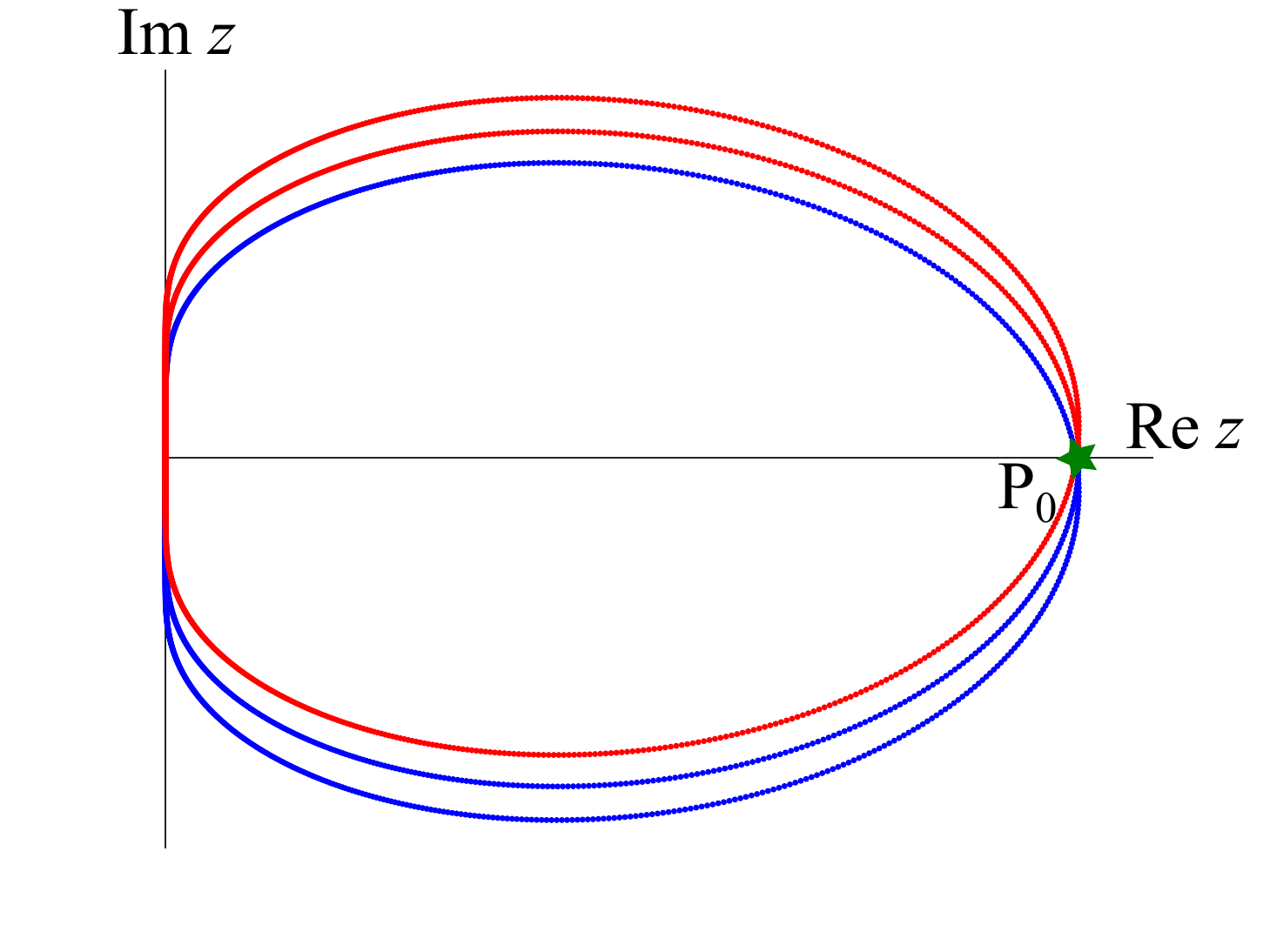}
\hfill
\textbf{e}\includegraphics[width=0.3\textwidth]{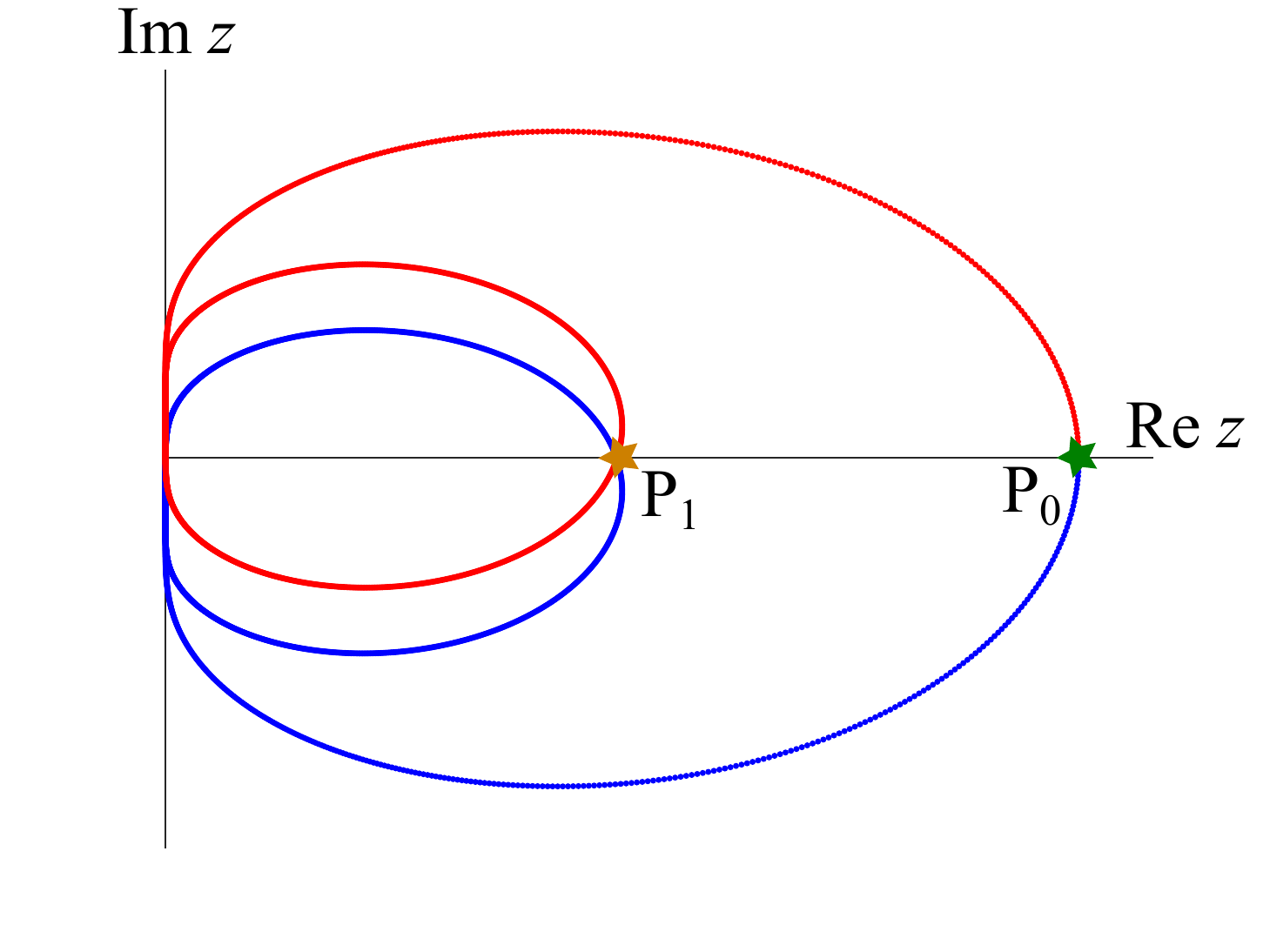}
\hfill
\textbf{f}\includegraphics[width = .31\textwidth]{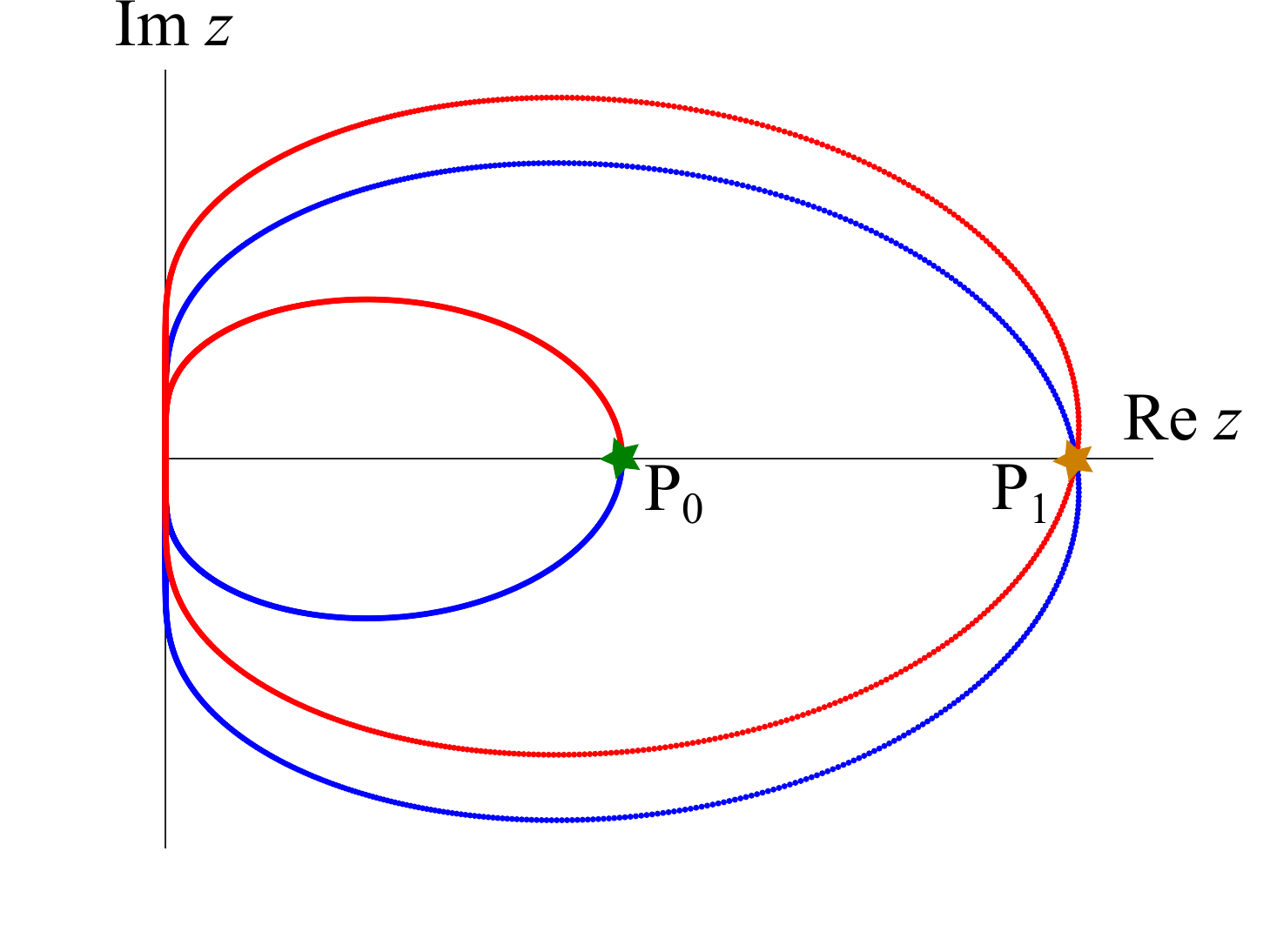}\\
\textbf{g}\includegraphics[width=0.3\textwidth]{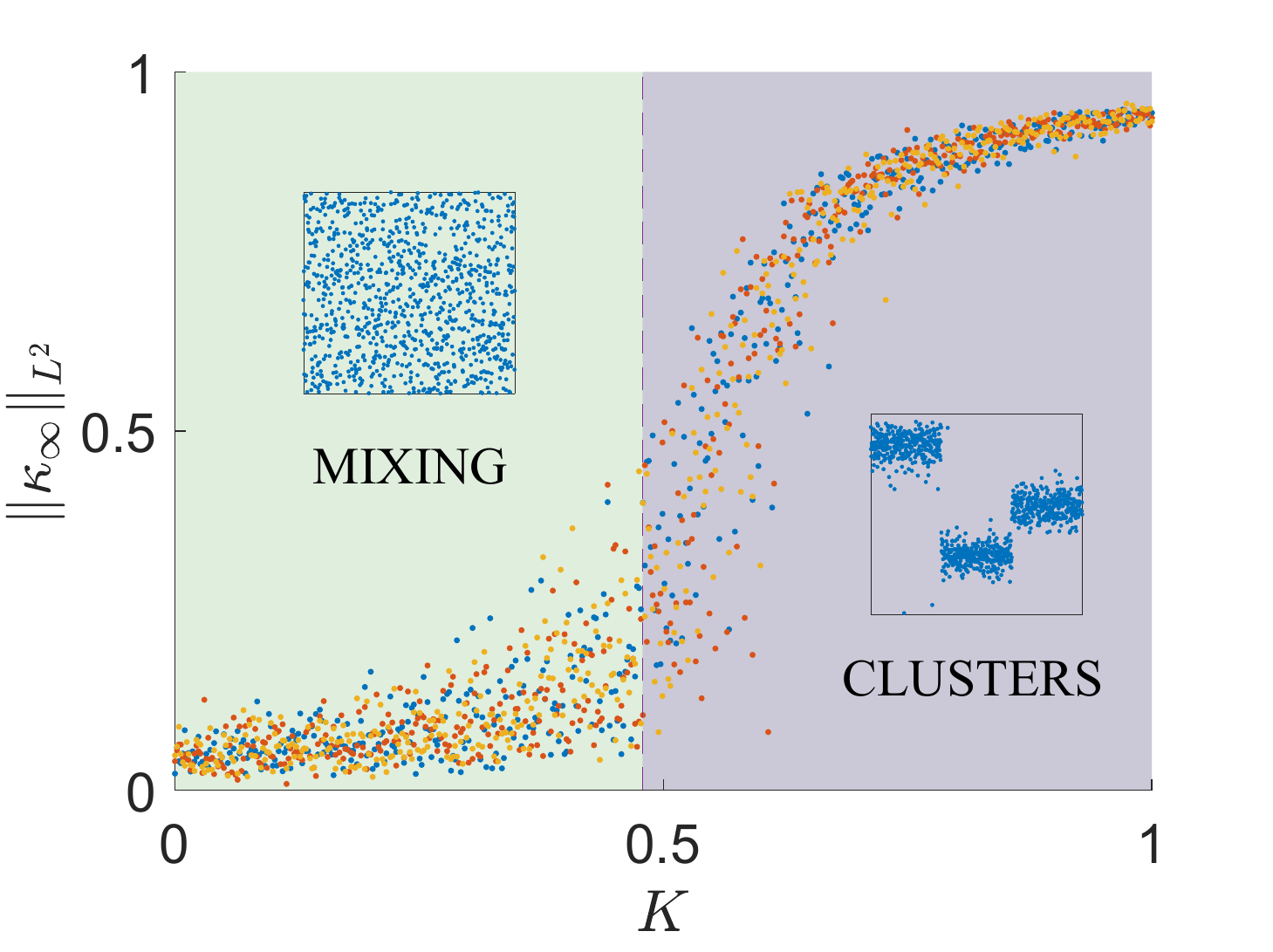}
\hfill
\textbf{h}\includegraphics[width=0.3\textwidth]{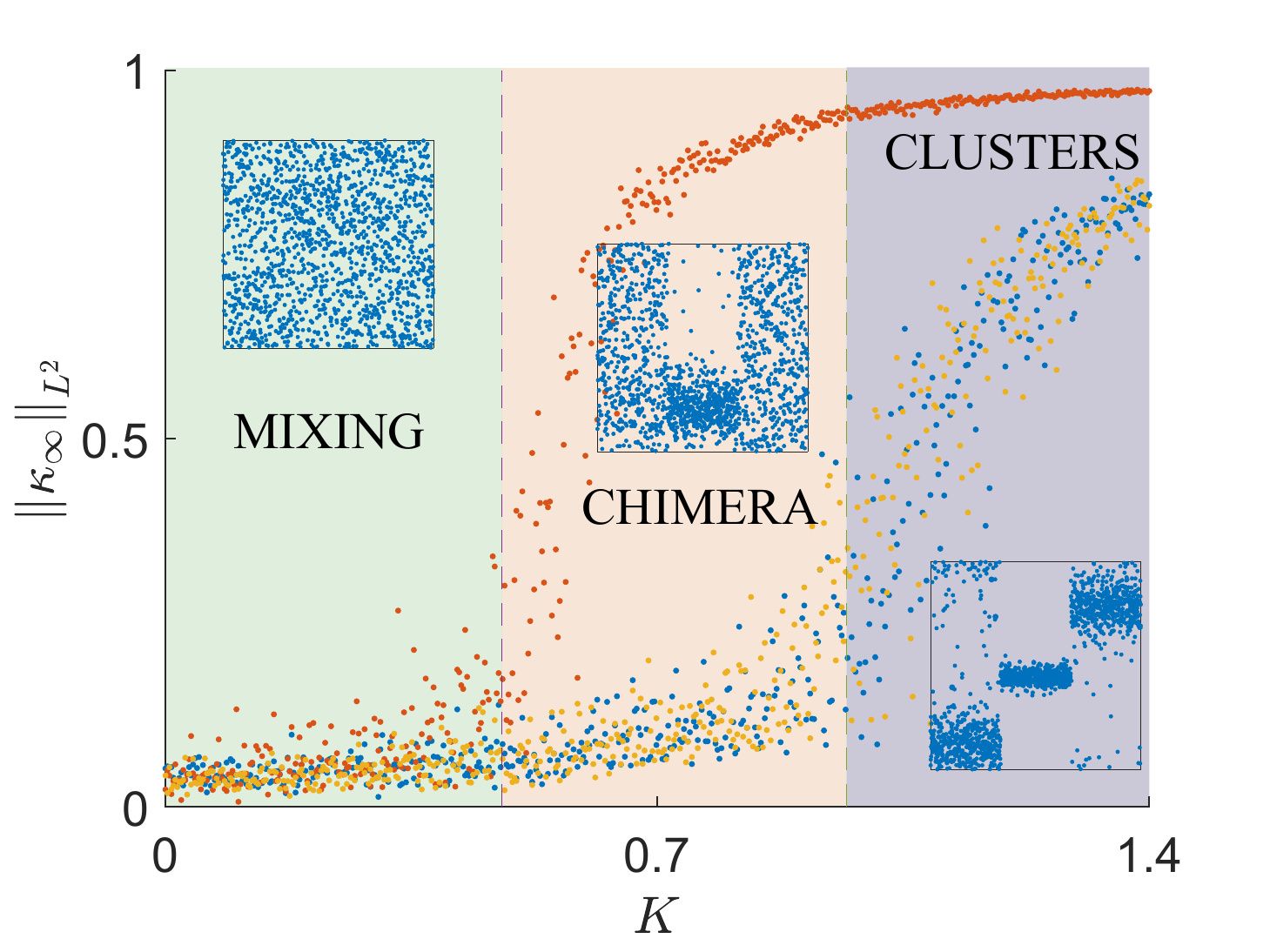}
\hfill
\textbf{i}\includegraphics[width = .31\textwidth]{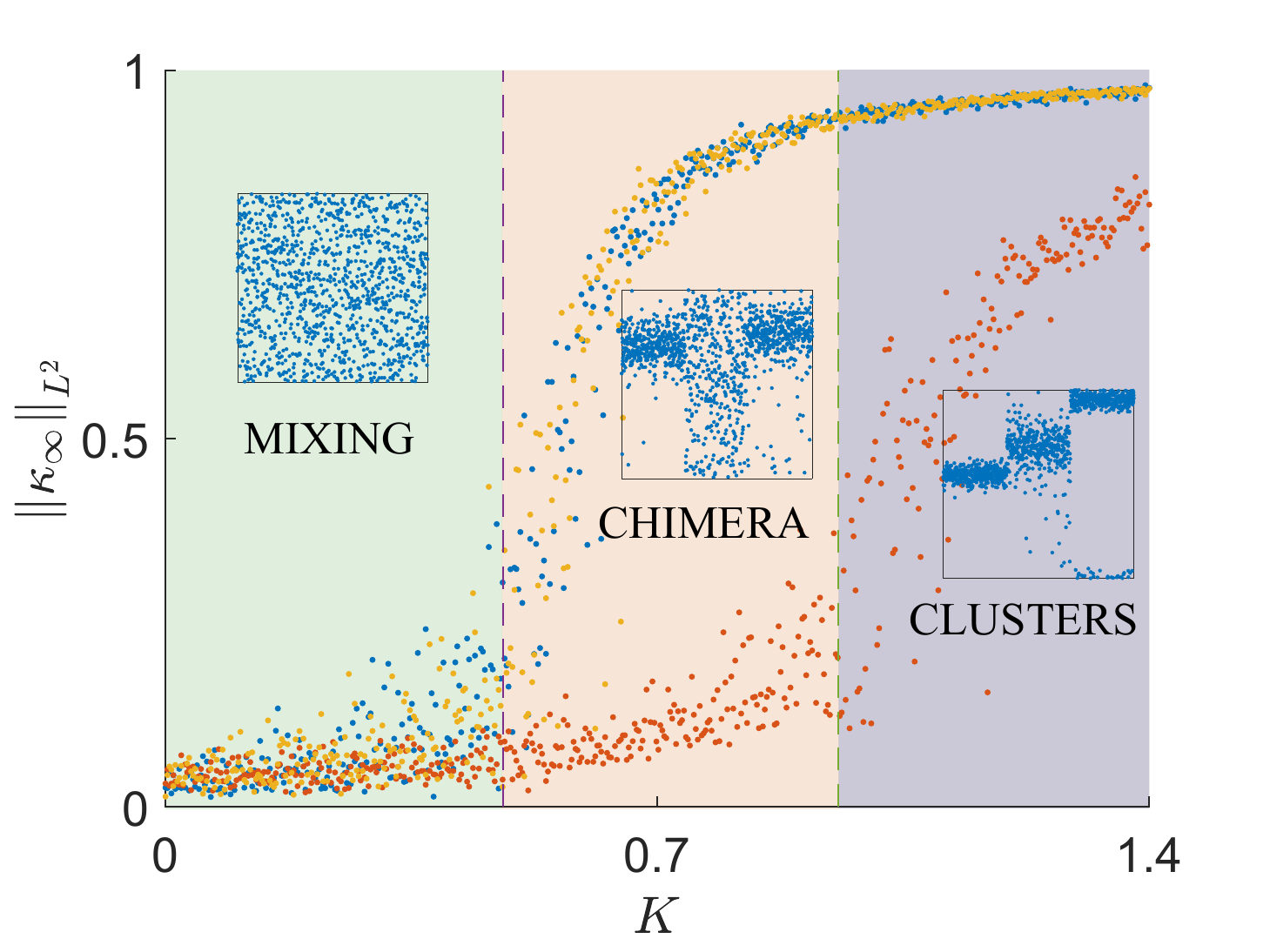} 
\end{minipage}
\caption{\textbf{a}-\textbf{c}) A family of trimodal probability distributions \eqref{phi-tri}.
  \textbf{d}-\textbf{f}) The corresponding critical curves. Under the deformation of the
  density functions, the point of triple intersection
  in \textbf{d} splits into a point of simple   intersection and a point of double intersection
   \textbf{e} and \textbf{f}.
   % The latter splits further into two points for simple intersection.
   Each point of simple intersection
   corresponds to a PF bifurcation producing a pattern with a stationary cluster, whereas a point of
  double intersection corresponds an AH bifurcation resulting in patterns with a pair of moving
  clusters. The bifurcation scenarios predicted by the diagrams in  \textbf{d}-\textbf{f} are
  shown in \textbf{g}-\textbf{i}. They are explained in the text.
}\label{f.tri}
\end{figure}

We are now prepared to discuss the bifurcations in the KM.
We start with the case of  an even unimodal $g$ (see Fig.~\ref{f.uni}\textbf{a}).
Although the bifurcation of mixing leading to the transition to synchrony for such $g$
is well understood (cf. \cite{StrMir91, Str00, Chi15a, Die16}), we use it  as an example to  explain
the Penrose's method. Below, we will apply this method to study bifurcations in a
families of bi- and tri- modal distributions (Figs.~\ref{f.bi}, \ref{f.tri}).

Using the symmetry of $g$, from \eqref{parC} we see that $\cC$ is symmetric about the $x$--axis.
It intersects the positive real semiaxis at a unique point $P_0=(x_0,0), x_0>0$.
Further, note that $G^{-1}(x_0)=0$ (Figure~\ref{f.bi}\textbf{a}, \textbf{b}).
From the $x$--equation in \eqref{parC} we find   $x_0=\pi g(0)$.
By the Argument Principle, the number of roots of \eqref{GK} in $\bbH^+$ is
equal to the winding number of $\cC$ about $K^{-1}$ \cite{Penrose60}.
Since for $K< K_c:=(\pi g(0))^{-1}$, $K^{-1}$ lies outside $\cC$ (Fig.~\ref{f.uni}\textbf{b}), and the winding
number is $0$. We conclude that for $K< K_c$, $\bT$ has no eigenvalues with positive real parts.
Thus, for $K<K_c$, mixing is linearly stable. In fact, it is asymptotically stable \cite[Theorem~4.1]{ChiMed19b}.
For $K>K_c$, on the other hand, the winding number is $1$.
As $K\to K_c+0$, $\lambda\to 0+$,  and at $K=K_c$, mixing undergoes a PF bifurcation. Using \eqref{v-lam}
and \eqref{v0-distribution}, we compute the eigenfunction corresponding to $\lambda=0$:
\begin{equation}\label{vPF}
  \upsilon_0(\omega, x) =\pi g(0)\delta_0(\omega)-\iu\mathcal{P}_0[g](\omega).
\end{equation}
Each term on the right--hand side of \eqref{vPF} has a singularity at $0$. The second
term also has a regular component. This determines the structure of the PLS bifurcating from
the mixing state (Fig.~\ref{f.uni}c). The delta function on  the right--hand side of \eqref{vPF} implies that 
the coherent cluster within the PLS is stationary. The regular component of  $\iu\mathcal{P}_0[g]$ yields the velocity
distribution within the incoherent group. The combination of these two terms yields the velocity
distribution within the PLS.

\subsection{Bimodal $g$}
\begin{figure}
  \centering
\begin{minipage}[b]{.98\textwidth}
\textbf{a}\includegraphics[width=0.3\textwidth]{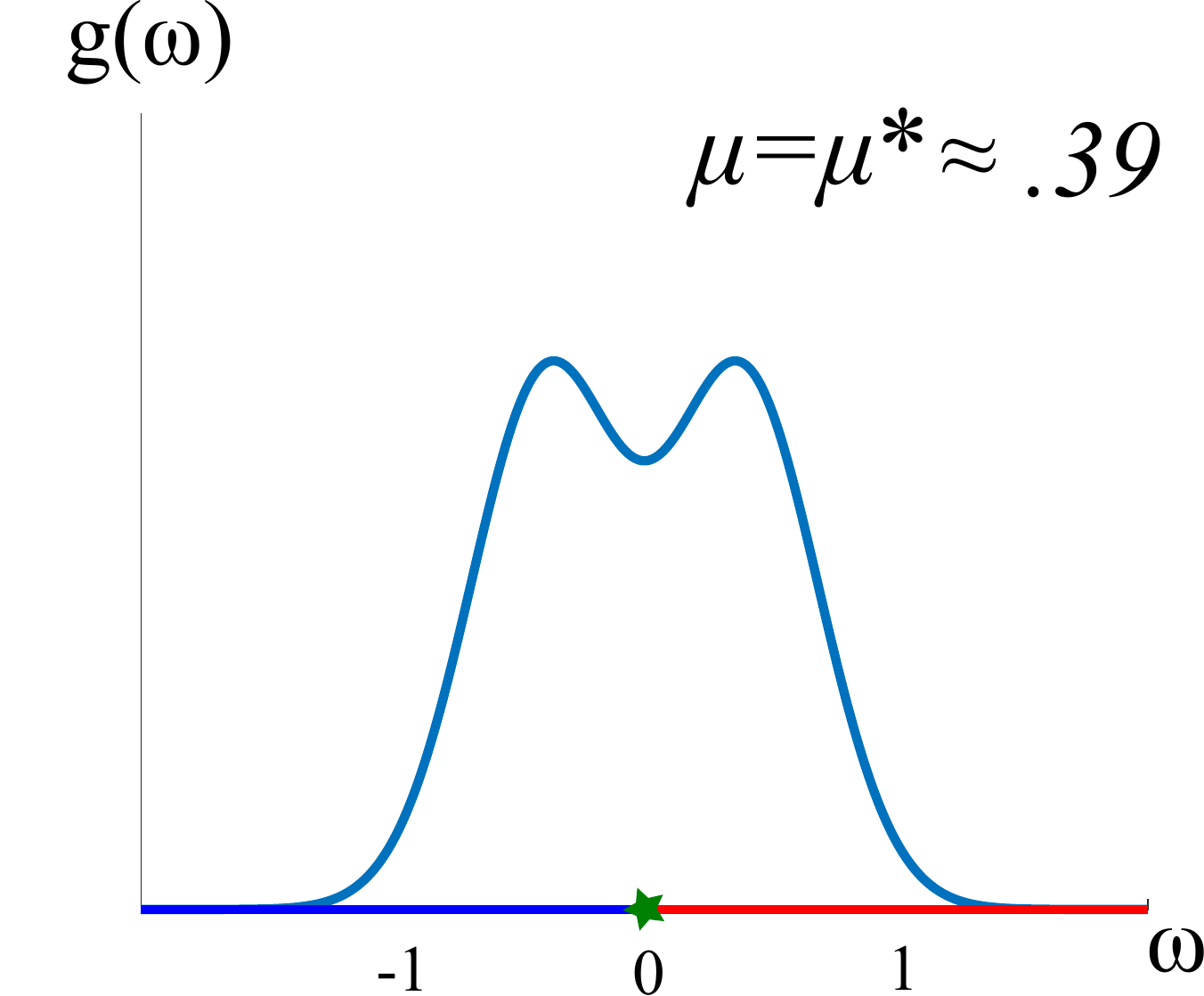}
\hfill
\textbf{b}\includegraphics[width=0.3\textwidth]{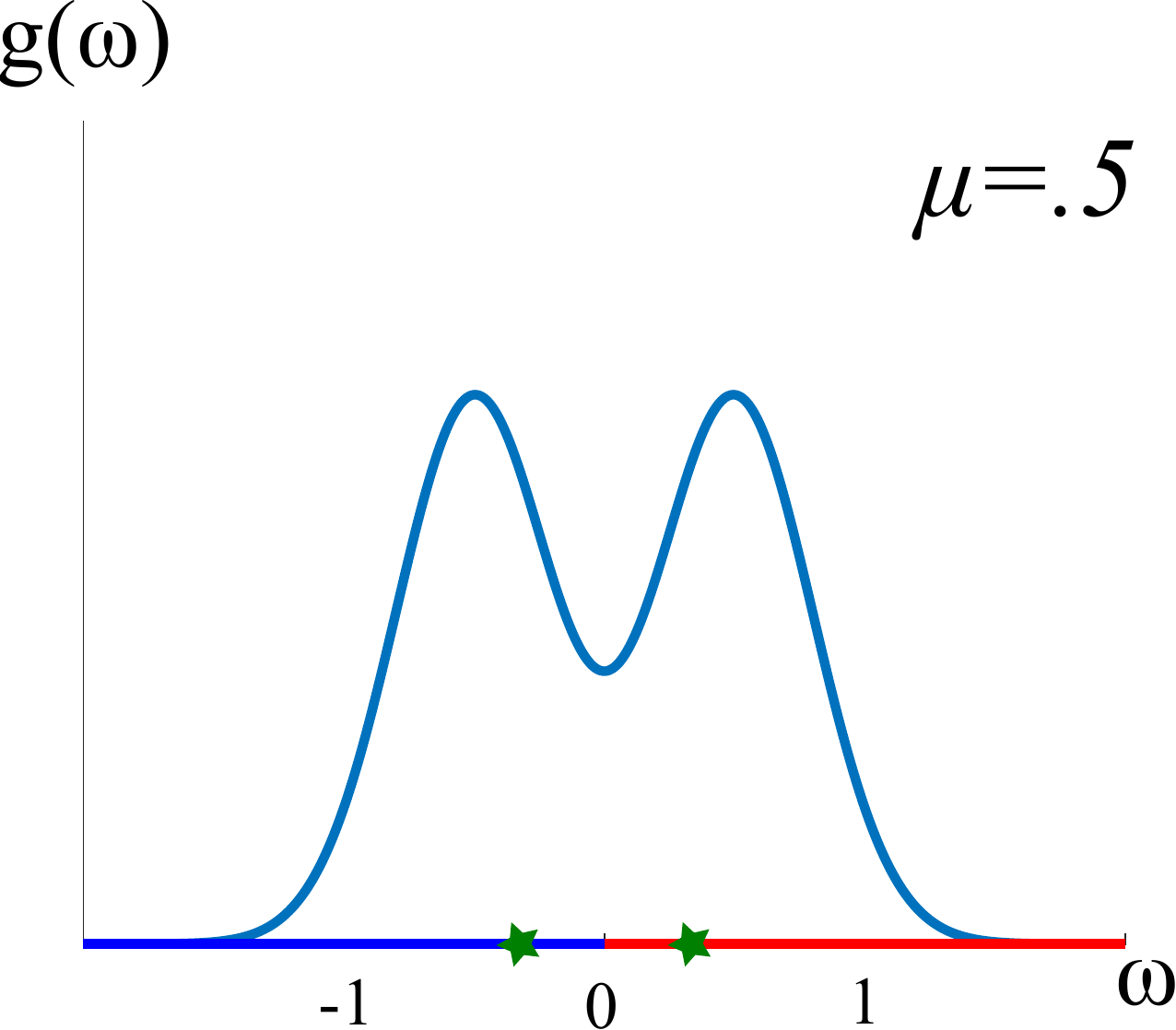}
\hfill
\textbf{c}\includegraphics[width=0.3\textwidth]{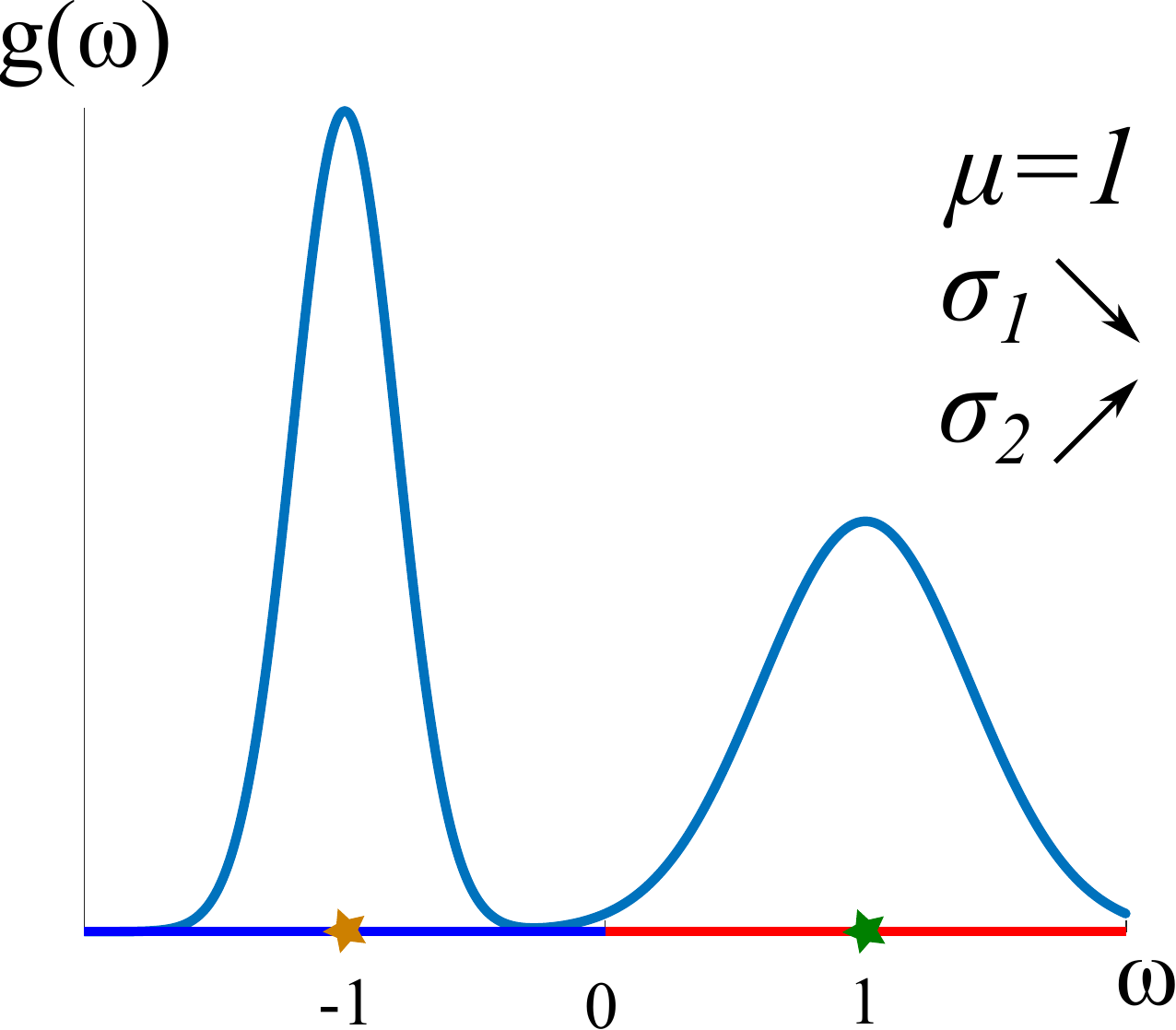}\\
\textbf{d}\includegraphics[width=0.3\textwidth]{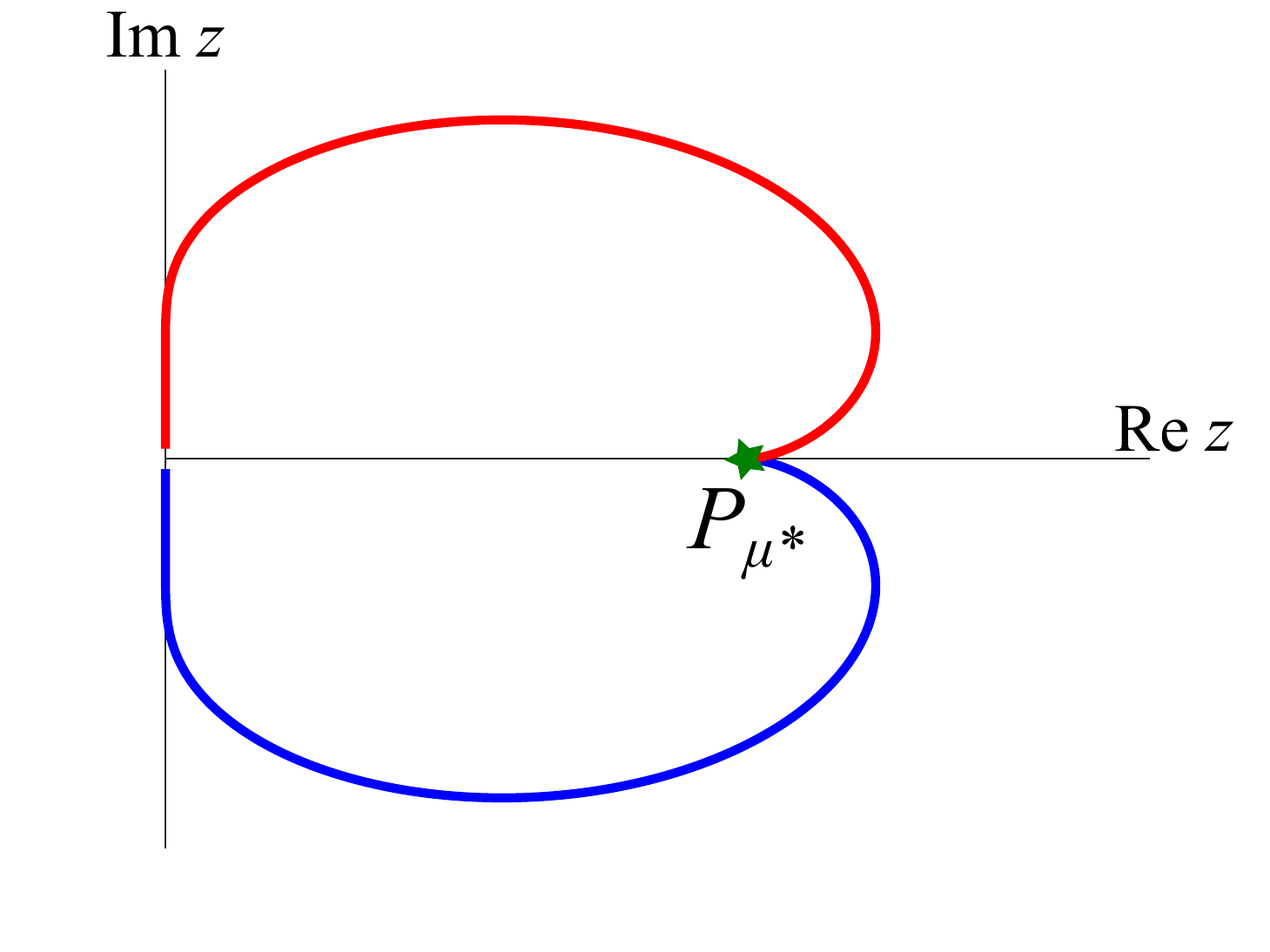}
\hfill
\textbf{e}\includegraphics[width=0.3\textwidth]{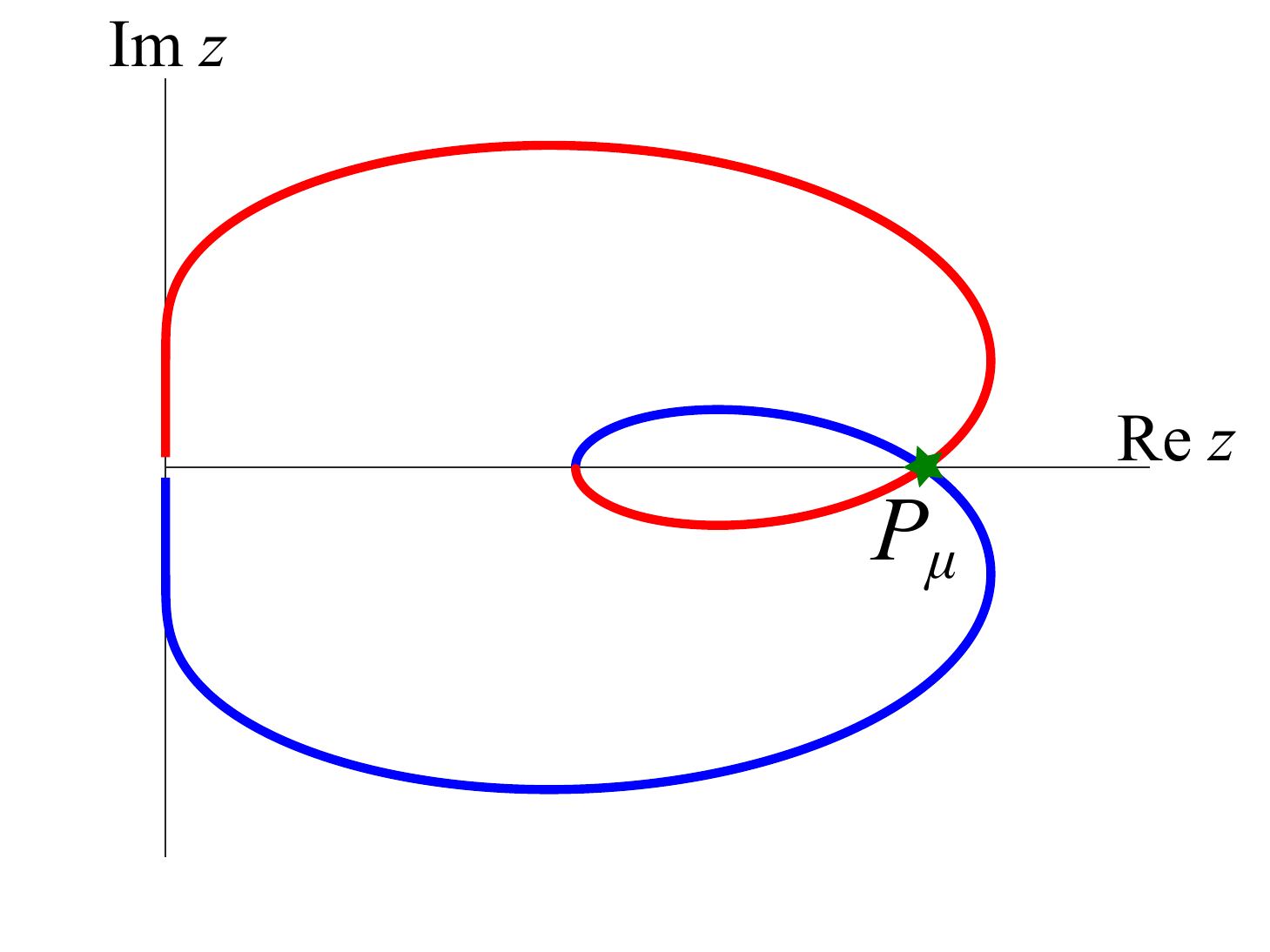}
\hfill
\textbf{f}\includegraphics[width=0.3\textwidth]{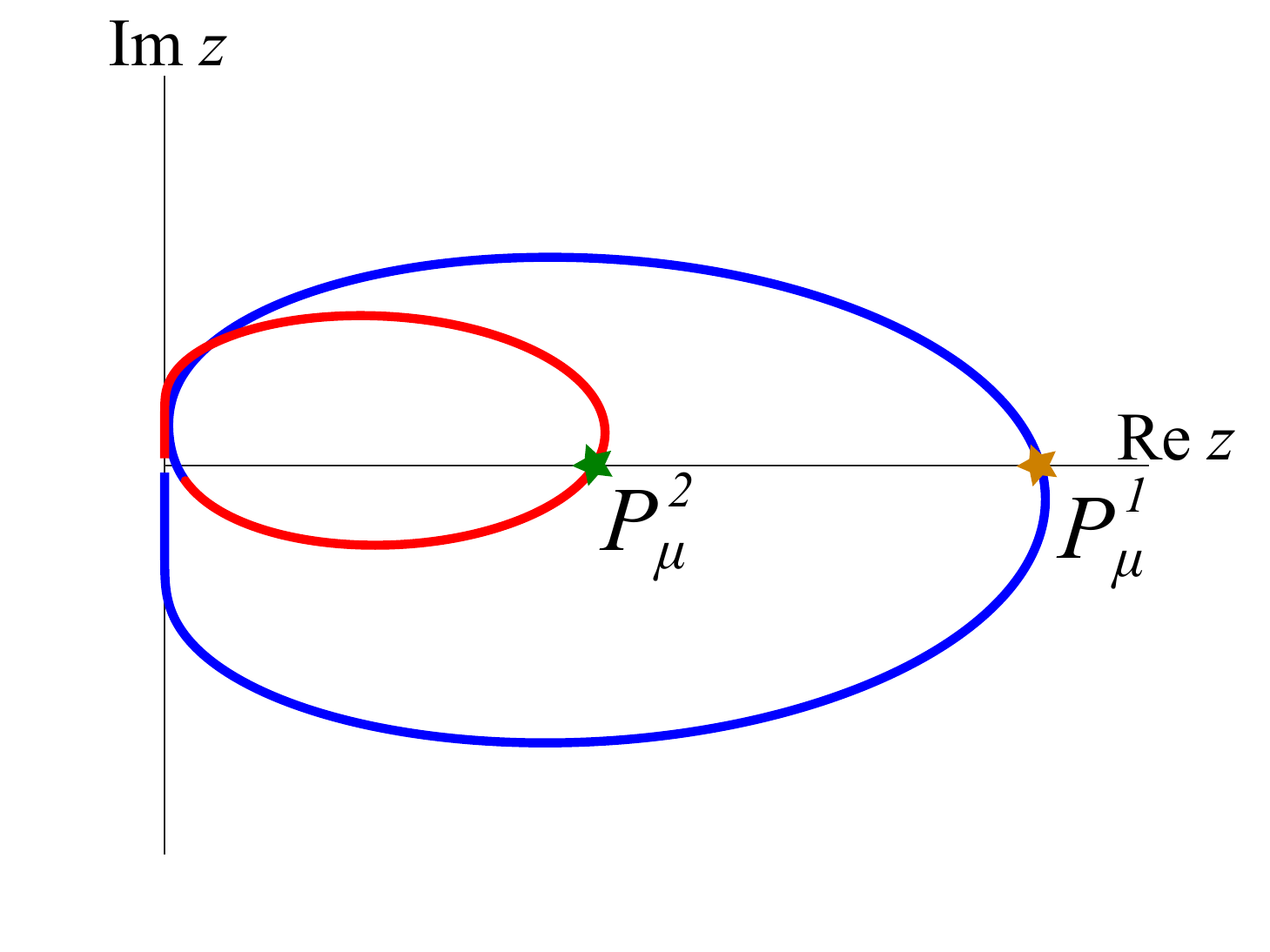}
\caption{\textbf{a}-\textbf{c}) Continuous deformation of the unimodal symmetric density $g$
  into a bimodal asymmetric one (\textbf{c}) and the plots of the corresponding
  critical curves (\textbf{d}-\textbf{f}). At the critical value $\mu=\mu^\ast$, $\cC_{\mu^\ast}$ develops
  a cusp (\textbf{d}). This corresponds to the codimension--$2$ bifurcation of mixing.
  The preimages of points of the intersection of the critical curve with the real
  axis $P_\mu$ and $P_\mu^{1,2}$ in (\textbf{d}-\textbf{f}) are indicated by stars in the corresponding plots in
  (\textbf{a}-\textbf{c}).
}\label{f.bi}
\end{minipage}
\end{figure}

\begin{figure}
  \centering
\begin{minipage}[b]{\textwidth}
\textbf{a}\includegraphics[width=0.32\textwidth]{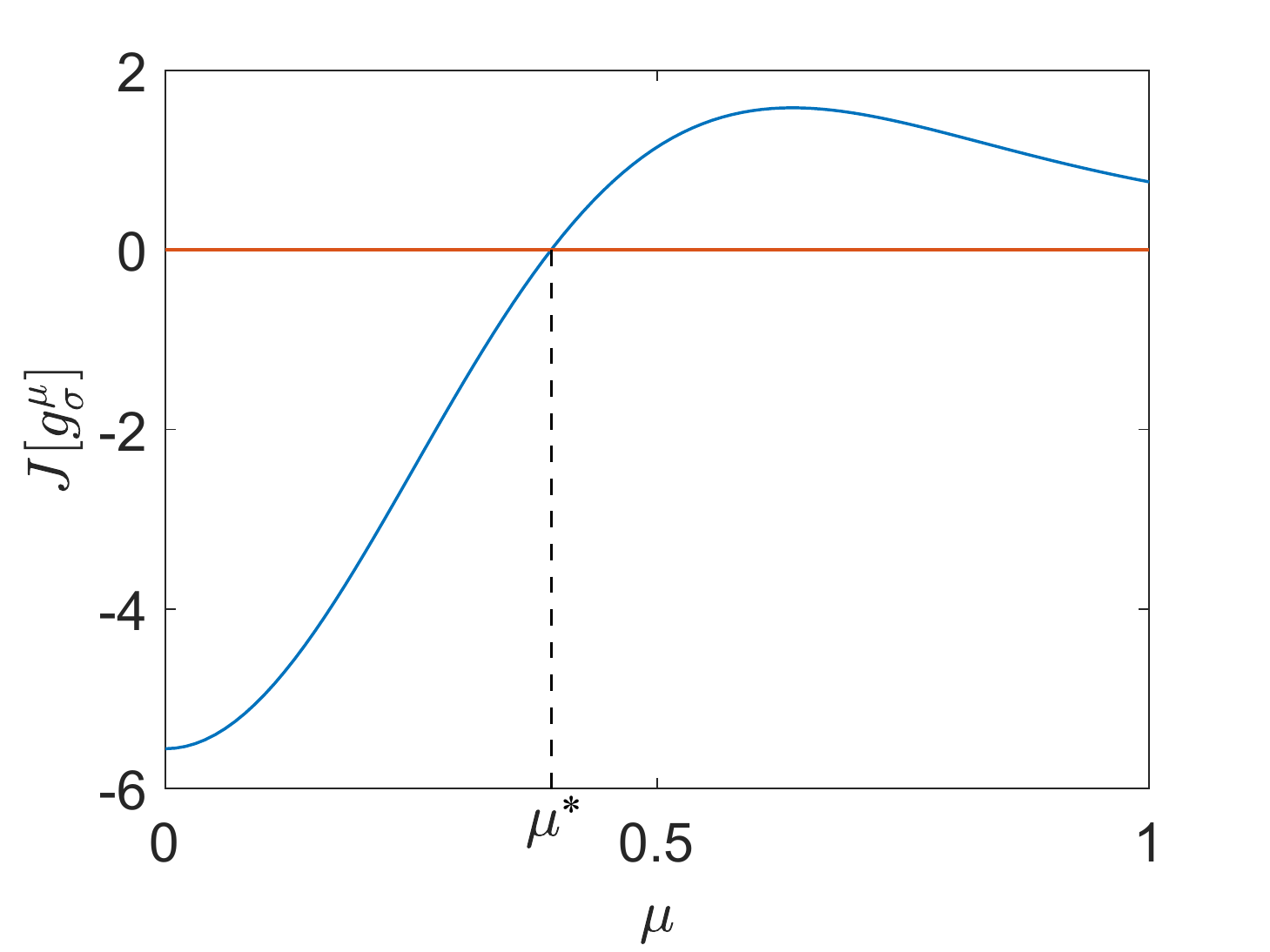}
\textbf{b}\includegraphics[width=0.29\textwidth]{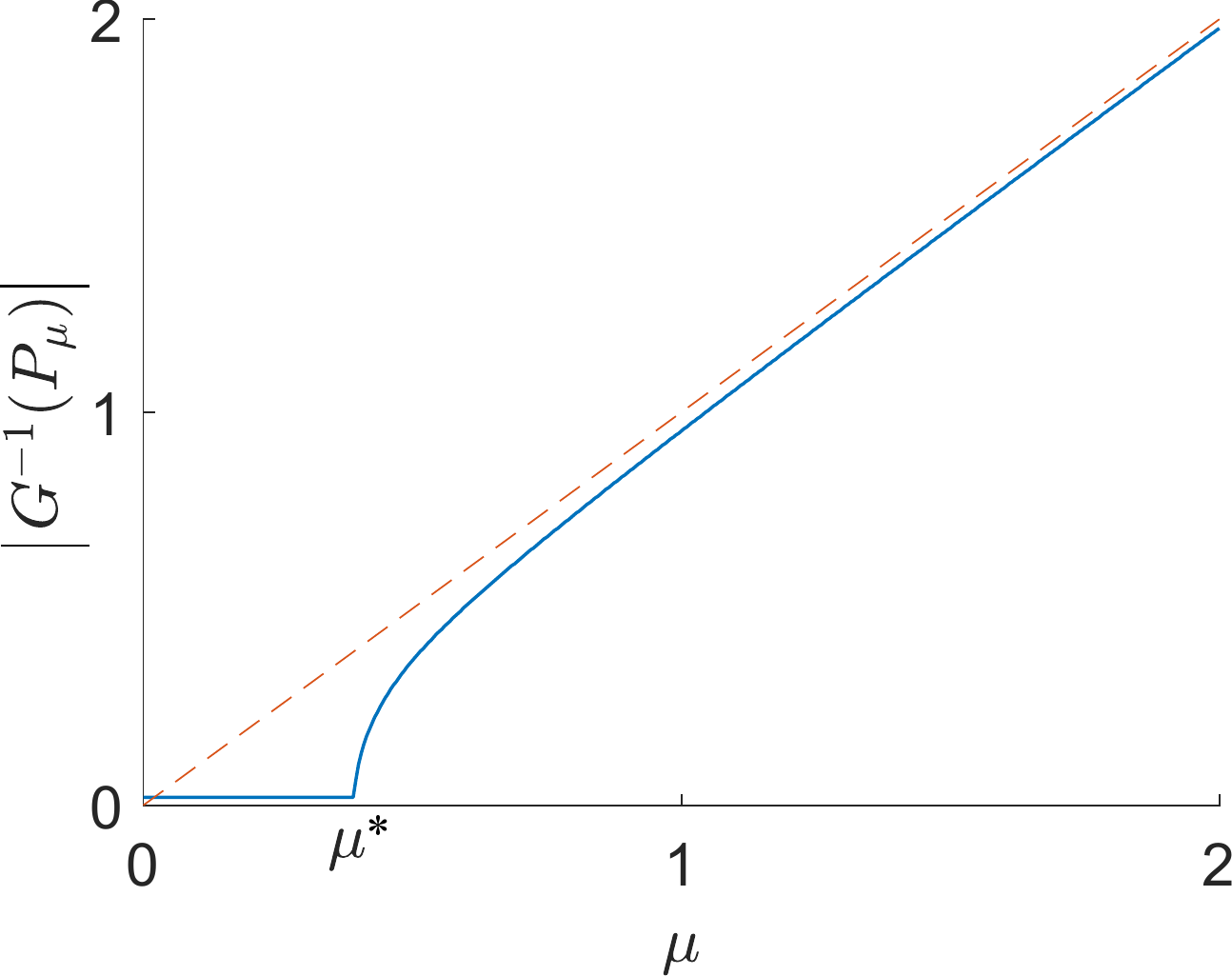}
\textbf{c}\includegraphics[width=0.33\textwidth]{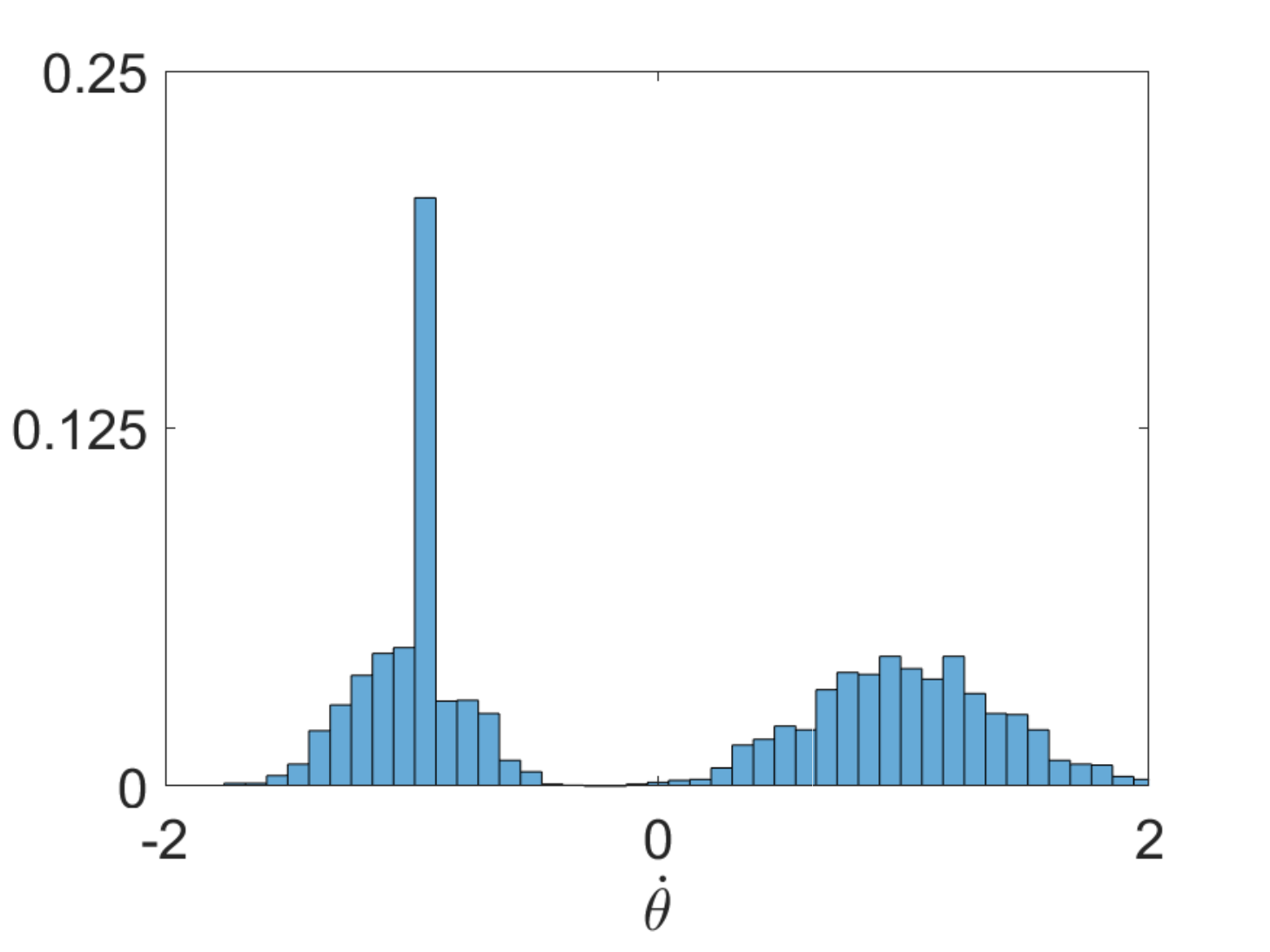}
\caption{\textbf{a}) The plot of $J[g_\sigma^\mu]$ vs $\mu$ (cf.~\eqref{cusp}). The zero of
  $J[g^\mu_\sigma]$ determines the critical value $\mu^\ast$.
  \textbf{b}) The plot of the absolute value of the two preimages $G^{-1}(P_\mu)$.
  Note that for $\mu>\mu^\ast$ outside a small neighborhood of $\mu^\ast$,
  $\left|G^{-1}(P_\mu)\right|\approx \mu,$ i.e., the two preimages of  $P_\mu$ lie near the peaks of the
  density $g$. \textbf{c}) The histogram of the
 velocity distribution
  within a chimera is fully determined by the singular  distribution $v_{\iu t_\mu+0}$
  (cf.~\eqref{chim-mode}).} \label{f.bi+}
\end{minipage}
\end{figure}

Next, we turn to the description of a bifurcation scenario connecting mixing to chimeras through $2$--clusters.
To this end, we continuously deform the unimodal distribution $g$ into a bimodal distribution,
preserving even symmetry as shown in Fig.~\ref{f.bi}\textbf{a},\textbf{b}. In our numerical experiments, we
use the following family of probability distribution functions:
\begin{equation}\label{phi}
  g^{\mu}_{\sigma_1,\sigma_2}(x)= \frac{1}{2\sqrt{2\pi}}
  \left\{ \frac{ e^{{-(x+\mu)^2\over 2\sigma_1^2}}} {\sigma_1} +
      \frac{e^{{-(x-\mu)^2\over 2\sigma_2^2}}}{\sigma_2 } 
      \right\}  
    \end{equation}
    When $\sigma_1=\sigma_2=:\sigma$, we collapse indices into one 
   $g^{\mu}_{\sigma}:=g^{\mu}_{\sigma,\sigma}$.
    First, we keep $\sigma_1=\sigma_2=:\sigma$ and increase $\mu$ from
    zero. We want to understand how the critical curve changes as $\mu$ is varied.
    The key events in the metamorphosis of $\cC$ are shown  Fig.~\ref{f.bi}\textbf{d}, \textbf{e}.
     % A qualitative analysis that follows
   %  will suffice for our purposes. More detailed can be derived using \eqref{parC}.
    % A more quantitative information
    % can be easily derived from \eqref{parC}.
    For small $\mu>0$, $\cC_\mu$ \footnote{From this point on, we 
      explicitly  indicate the dependence of $\cC,$ $x,$ and $P$ on $\mu$.}
    is diffeomorphic to $\cC_0$ in a neighborhood
    of $P_0$, the point the intersection of $\cC_0$ with the real axis.
    % Here, we used the subscript
    % to indicate the dependence on $\mu$ explicitly.
    At a critical value $\mu^\ast>0$,
    $\cC_{\mu^\ast}$ develops a cusp at $P_{\mu^\ast}$ (see Fig.~\ref{f.bi}\textbf{d}). To identify the condition
    for the cusp, we look for the value of $\mu$, at which the condition of the Inverse Function
    Theorem fails for $G$. By \eqref{parC} this occurs when $\left. dy/dt\right|_{t=0}=0$, i.e., 
    \begin{equation}\label{cusp}
      J[g^{\mu^\ast}_\sigma]:=\int_0^\infty \frac{ (g^{\mu^\ast}_\sigma)^\prime(s)}{s}ds=0
      \end{equation}
      (see Fig.~\ref{f.bi+}\textbf{a}).

    For $\mu>\mu^\ast$ there is a point on the real axis $P_\mu$, which has
    two preimages under $G$: $\pm \iu\nu$ (Fig.~\ref{f.bi}\textbf{b}, \textbf{e}).
    Thus, for $\mu>\mu^\ast$ mixing loses stability
    through the  AH bifurcation and not through the PF bifurcation. At the AH bifurcation,
    $\bT$ has a pair of complex conjugate eigenvalues $\pm \iu \nu$. 
The corresponding eigenfunctions are given by \eqref{v-lam}, \eqref{viy-distribution}
\begin{equation}\label{AHmode}
  v_{\pm \iu \nu+0}=\pi g^\mu_\sigma(\pm \nu)\delta_{\pm \nu}-
  \iu\mathcal{P}_{\pm\nu}[g_\sigma^\mu].
\end{equation}

The first term on the 
right-hand side of  $v_{\iu\nu+0}$ (cf.~\eqref{AHmode}) is localized at $\omega=\nu$.
The second term is 
singular at $\nu$ too, but also has a regular component. The combination of
of $v_{\iu\nu+0}$ and $v_{-\iu\nu+0}$
results in splitting the population into two groups of approximately equal
size rotating with the velocities centered around $\pm\nu$. Thus, for
$\mu>\mu^\ast$ mixing bifurcates into a $2$--cluster state. At $\mu=\mu^\ast$, where
the regions of the PF and AH bifurcations meet, we have a codimension--$2$ bifurcation, whose unfolding
contains the transitions to synchronization, $2$--clusters, and to chimeras, as we are going to see
next.

\begin{figure*}
  \centering
  \begin{minipage}[b]{.98\textwidth}
    \textbf{a}\includegraphics[width=0.48\textwidth]{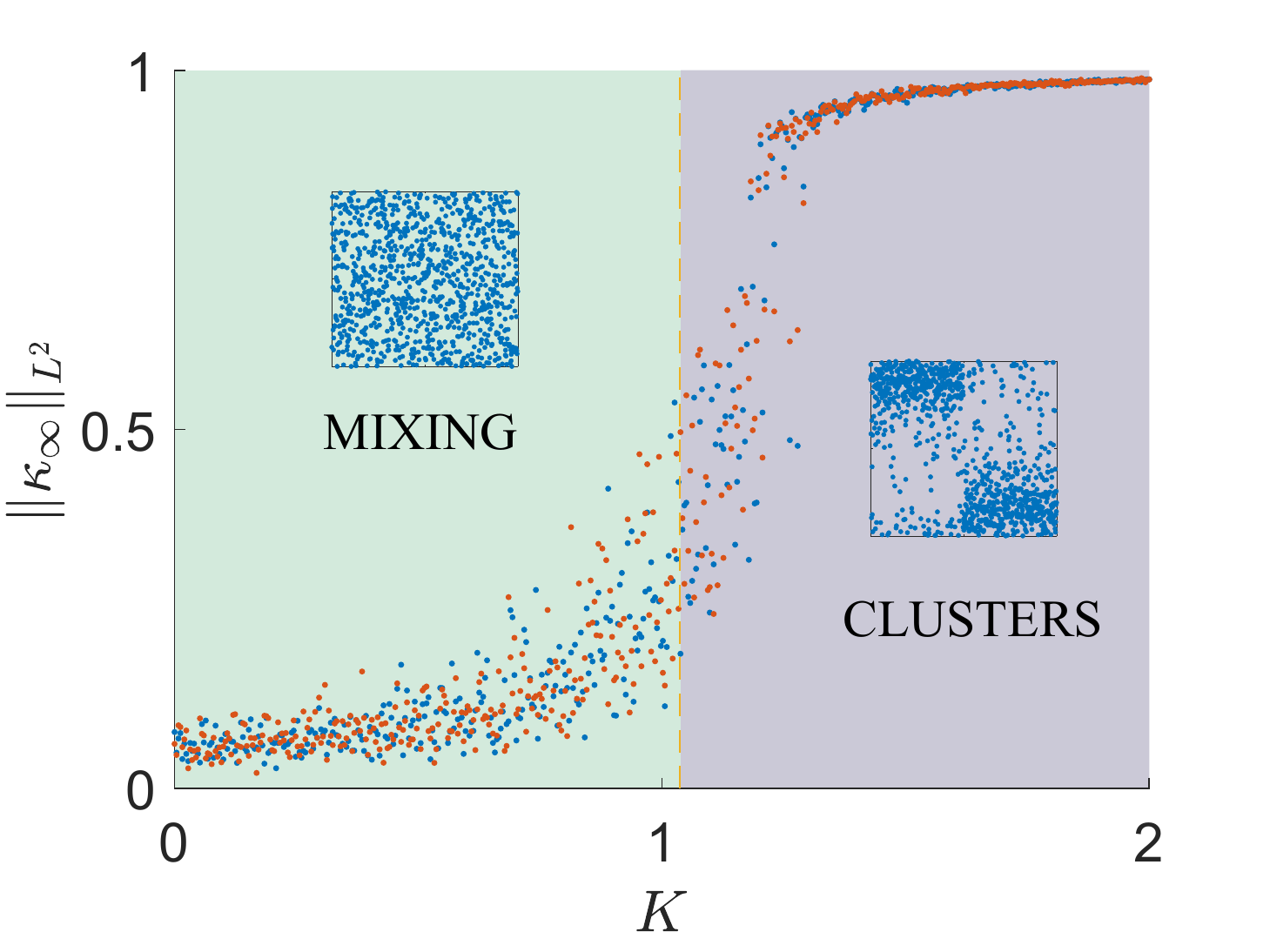}\hfill
\textbf{b}\includegraphics[width=0.48\textwidth]{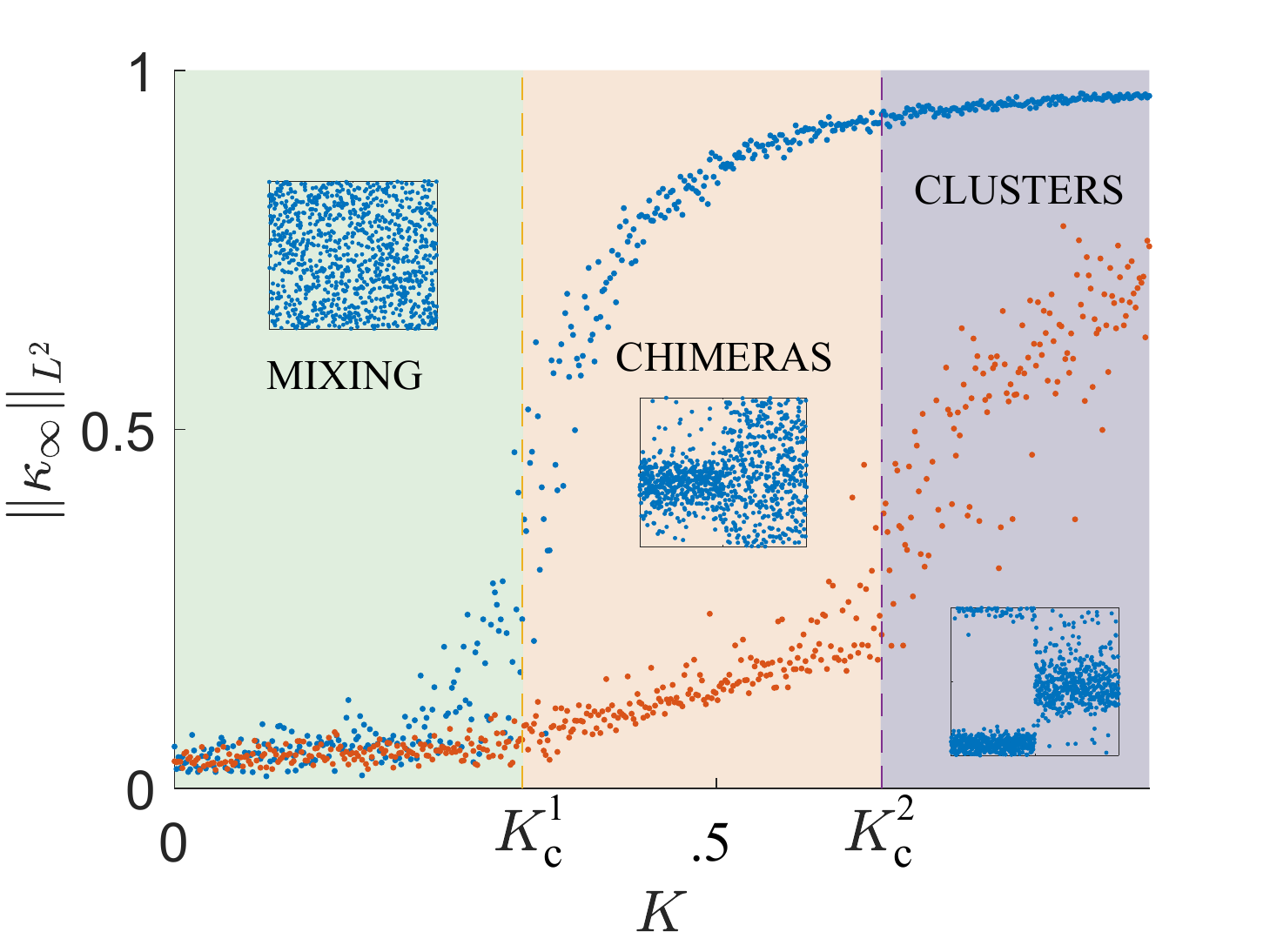}
% \textbf{b}\includegraphics[width=0.48\textwidth]{./FF2/nn_bif_diag.pdf}
\caption{The bifurcation diagrams corresponding to the symmetric  and asymmetric bimodal distributions
  for all--to--all (\textbf{a} and \textbf{b}, respectively). Colored dots indicate the value of the order
  parameter computed for each cluster separately for different values of $K$ and different realizations
  of $\omega_i$'s. Dots of different colors represent different clusters.
  In (\textbf{a}), the loss of stability of mixing
  results in the creation of a $2$--cluster state.
In (\textbf{b}),
  a chimera is born at the loss of stability of mixing at $K_c^1$. It bifurcates into a moving
  $2$--cluster  at $K_c^2$. Here,  $\kappa_\infty(x)$ stands for the asymptotic value of the
  local order parameter $\kappa (t,x)$ (cf.~\eqref{corder}). Note that the bifurcations at
  $K_c^1$ and $K_c^2$ affect clusters practically separately.
  }\label{f.bi-bif}
\end{minipage}
\end{figure*}   

We now fix $\mu>\mu^\ast$ and break the even symmetry of $g^\mu_\sigma$ by
decreasing $\sigma_1$ and increasing $\sigma_2$ (see Fig.~\ref{f.bi}\textbf{c}).
This affects the critical curve $\cC_{\mu,\sigma_1,\sigma_2}$ in the following way.
The point of double intersection $P_\mu$ splits into two points of intersection with the
real axis: $P_\mu^1=(x_\mu^1, 0)$ and $P_\mu^2=(x_\mu^2, 0)$ with $0<x_\mu^2<x_\mu^1$
(see Fig.~\ref{f.bi}\textbf{f}).
Note that the preimages of these points under $G$ are still very close to the maxima of
$g^\mu_{\sigma_1,\sigma_2}$ (see Fig.~\ref{f.bi}\textbf{c} and Fig.~\ref{f.bi+}{\bf b}). In particular, the preimage of $P_\mu^1$
is approximately $-\iu\mu$, the center of the more localized peak of $g^\mu_{\sigma_1,\sigma_2}$ .
This implies that mixing loses stability at $K_c^1\approx (\pi g^{\mu}_{\sigma_1,\sigma_2}(-\mu))^{-1}$.
The bifurcating eigenvalue
$\lambda=\iu\nu_1 (\nu_1\approx -\mu)$ and the corresponding eigenfunction
\begin{equation}\label{chim-mode}
  v_{\iu\nu_1+0} = \pi g^{\mu}_{\sigma_1,\sigma_2}(\nu_1) \delta_{\nu_1} -
  \iu \mathcal{P}_{\nu_1}[g^{\mu}_{\sigma_1,\sigma_2}].
\end{equation}
Note that the first term on the right hand side of \eqref{chim-mode} is a singular distribution
localized at $\nu_1$. The second term  has a singularity at $\nu_1$, but its regular part  has some
'weight' near $\nu_2\approx \mu$. These features translate into the velocity distribution within a chimera:
there is a tightly localized peak around $-\mu$ (the coherent group) and a broader
peak near $\mu$ (the incoherent group) (Fig.~\ref{f.bi+}\textbf{c}).

From the Penrose diagram in Fig.~\ref{f.bi}\textbf{f} we can read off the region of existence of chimeras.
Recall that $(x_\mu^{1,2},0)$ are the points of intersection of $C_\mu$ with the real axis (see Fig.~\ref{f.bi}\textbf{f}).
For  $K>K^2_c$, the winding number
of $\cC_\mu$ about $K^{-1}$ is $2$. The second unstable mode is given by
\begin{equation}\label{2nd-mode}
  v_{\iu\nu_2+0} = \pi g^{\mu}_{\sigma_1,\sigma_2}(\nu_2) \delta_{\nu_2} -
  \iu \mathcal{P}_{\nu_2}[g^{\mu}_{\sigma_1,\sigma_2}].
\end{equation}
Recall that $\nu_2\approx \mu$. The first term on the right hand
side of \eqref{2nd-mode} indicates the formation of a coherent cluster moving with velocity approximately
equal to $\mu$. Thus, chimera state is transformed into a pair of clusters moving with the speed $\mu$ in
opposite directions. We conclude that the chimera state is born at
$K^1_c\approx \left(\pi g^\mu_{\sigma_1,\sigma_2}(-\mu)\right)^{-1}$,
when mixing loses stability. This corresponds to the first point of intersection of the
critical curve with the real axis $P^1_\mu$ (Fig.~\ref{f.bi}\textbf{f}). The chimera state dissappears at
$K^2_c\approx \left(\pi g^\mu_{\sigma_1,\sigma_2}(\mu)\right)^{-1}$, when the second unstable mode is created. 
This corresponds to the second intersection point $P^2_\mu$ (Fig.~\ref{f.bi}\textbf{f}).
The existence region $(K_c^1, K_c^2)$, predicted by the Penrose diagram, agrees extremely well
with numerical simulations (Fig.~\ref{f.bi-bif}\textbf{b}). Moreover, the bifurcations at $K_c^1$ and
$K_c^2$ affect the coherent and incoherent clusters practically separately. This provides a general mechanism
for formation of chimera states. Under more restrictive assumptions a related
mechanism of creating chimeras by controlling fluctuations in separate clusters
for the KM with inertia was presented in \cite{MM21}.

There is an important distinction between the primary and secondary bifurcations corresponding
to points $P_\mu^1$ and $P_\mu^2$ in the diagram shown in Figure~\ref{f.bi}\textbf{f}.
The former signals the appearance of the postive eigenvalue in the spectrum of $\bT$,
the operator linearized about mixing. This corresponds to the PF bifurcation of mixing at
$K=K_c^1$ (Fig.~\ref{f.bi-bif}\textbf{b}). The meaning of $P_\mu^2$ is different. It corresponds
to the emergence of the second positive eigenvalue of $\bT$ at $K=K_c^2$. Apparently, the linearization
about mixing is still relevant for values of $K$ near $K_c^2$, as the two unstable modes of $\bT$ capture the
transformations in the system dynamics near $K_c^2$, and $(K_c^1, K_c^2)$ provides a good 
estimate of the region of existence of chimera states (Fig.~\ref{f.bi-bif}\textbf{b}). The secondary
bifurcation at $K_c^2$ may not be a bifurcation in the strict sense of the word,
but it is useful for interpretting the spatial
patterns in the KM  (Fig.~\ref{f.bi-bif}\textbf{b}), as it marks the end
of the region for chimeras.

\begin{figure}
   \centering
\begin{minipage}[b]{.98\textwidth}
  \textbf{a}\includegraphics[width=0.3\textwidth]{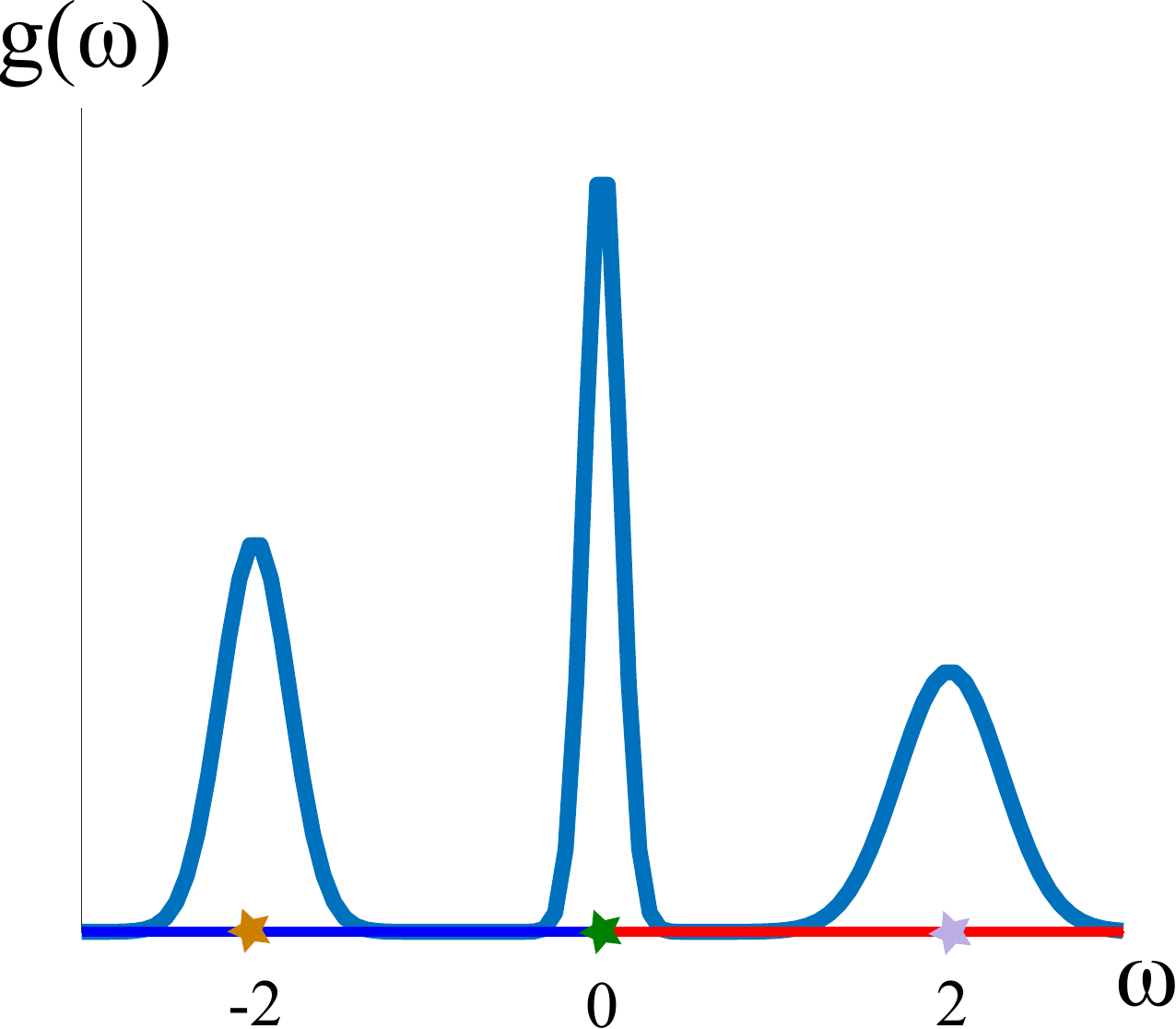}
\hfill
\textbf{b}\includegraphics[width=0.3\textwidth]{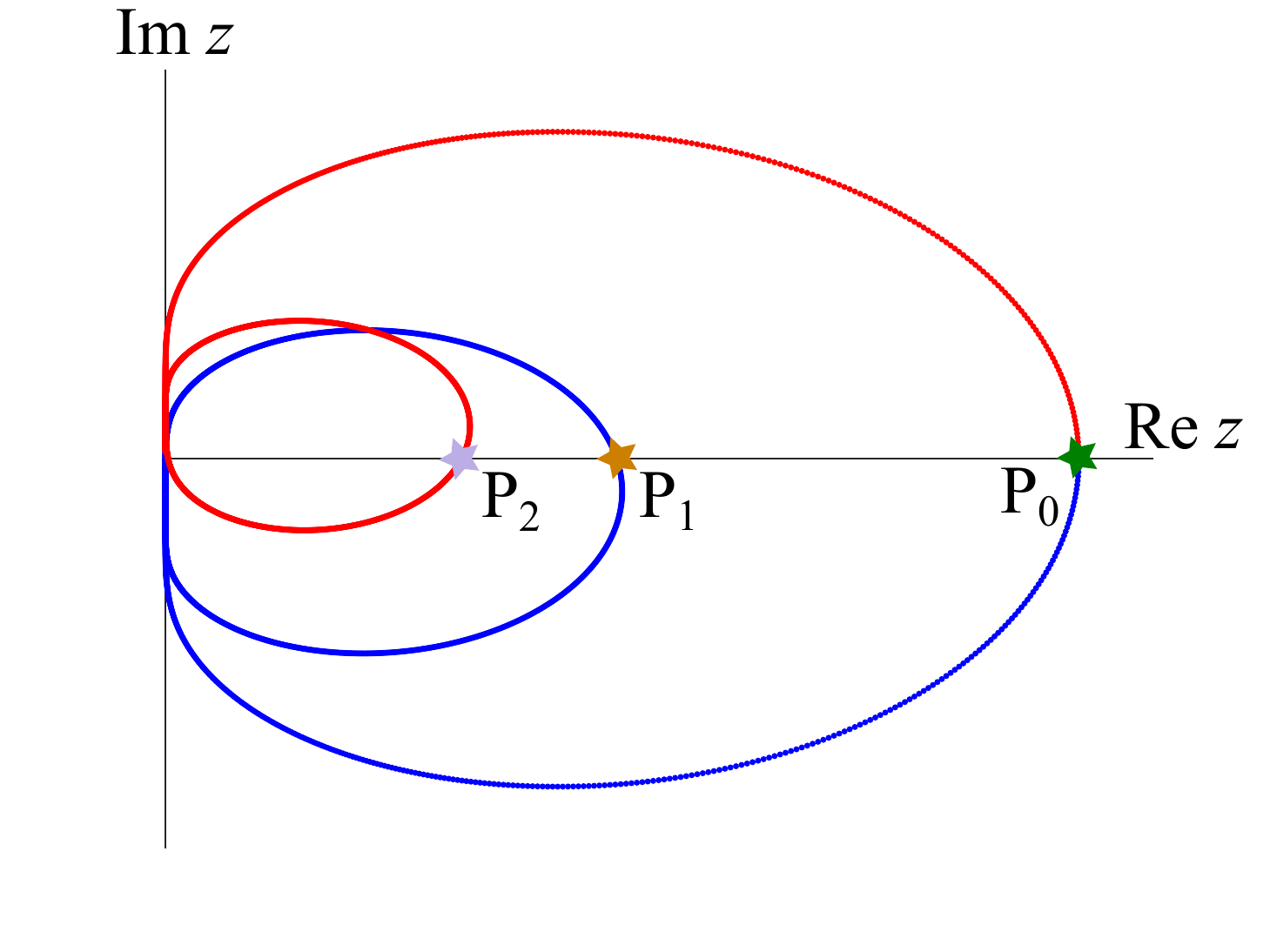}
\hfill
\textbf{c}\includegraphics[width=0.3\textwidth]{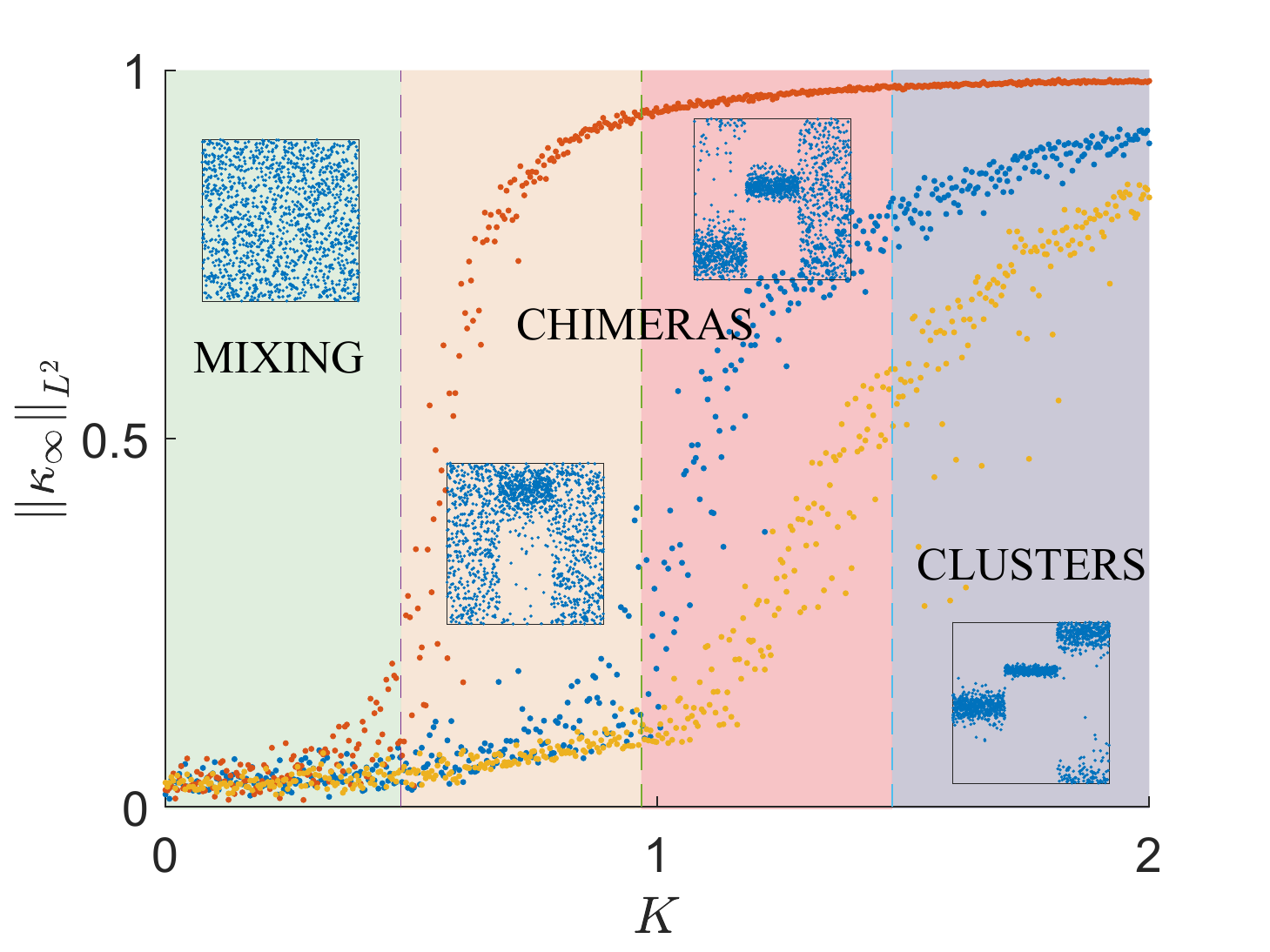}
\end{minipage}
\caption{Symmetry breaking  in the trimodal distribution shown
  in Figure~\ref{f.tri}\textbf{b} results in splitting of the
  point of the double intersection into two simple intersection points.
  Thus, the Penrose diagram predicts a PF bifurcation and two secondary
  bifurcations. The former
  produces a chimera with the stationary coherent cluster in the middle (\textbf{c}, beige),
  the first secondary bifurcation results in  a chimera with two coherent clusters
  (a stationary cluster  in the middle and a traveling one  on the left)
  (\textbf{c}, pink), and the last secondary bifurcation  produces a three-cluster state
  with the stationary middle cluster
  and two clusters rotating in opposite directions (\textbf{c}, purple).
}\label{f.tri+}
\end{figure}

\subsection{Trimodal $g$}
To give a more complete picture of possible bifurcation scenarios in the KM with multimodal
distributions of the intrinsic frequencies, we discuss bifurcations in the KM with trimodal family  
of distributions. In the numerical experiments used in this section,  we take the probability distribution
functions of the following form:
\begin{equation}\label{phi-tri}
	g^\mu_{\sigma_1,\sigma_2,\sigma_3} (x) = \frac{1}{3\sqrt{2\pi}} \left\{\frac{ e^{{-(x+\mu)^2\over 2\sigma_1^2}}} {\sigma_1} +\frac{ e^{{-x^2\over 2\sigma_2^2}}} {\sigma_2}+
		\frac{e^{{-(x-\mu)^2\over 2\sigma_3^2}}}{\sigma_3 }  \right\}.
            \end{equation}

            To study bifurcations in the model with trimodal frequency distribution we employ the same
           strategy as above. We first locate the highest codimension master bifurcation of mixing,
            whose unfolding contains all principal bifurcation scenarios. To this end, we fix $\mu>0$
            and choose $\sigma_1,\sigma_2,$ $\sigma_3$ so that the critical curve $\mathcal{C}$ has
            a point of triple intersection with the real axis $P_0$ (see Fig.~\ref{f.tri}\textbf{d}).
            In our numerical simulations, we used $\mu=2$ and $\sigma_{2} = 0.2$, and
            $\sigma_1=\sigma_3\approx 0.1976$, i.e., all peaks are practically the same (see Fig.~\ref{f.tri}\textbf{a}).
 The intersection point $P_0$ has three preimages under $G$: $G^{-1}(P_0) \approx \{\pm 2i, 0\}$.
            Thus, the loss of stability of mixing takes place through a PF-AH bifurcation. The
            diagram in Fig.~\ref{f.tri}\textbf{g} shows the bifurcation of mixing
            producing a $3$-cluster state. The middle cluster is stationary as implied by the PF bifurcation and
            the two outer clusters are moving with opposite velocities as implied by the AH bifurcation.

            Next, we deform the distribution in Figure~\ref{f.tri}\textbf{a} reserving even symmetry in two different ways.
            First, we increase $\sigma_1$ and $\sigma_3$ keeping them equal (see Figure~\ref{f.tri}\textbf{b}).
            Under this deformation, the triple intersection point splits into  a simple intersection point $P_0$ and a point
            of double intersection $P_1$ (see Figure~\ref{f.tri}\textbf{e}). This results in a PF bifurcation followed
            by the AH bifurcation. The former produces a chimera state with a stationary middle cluster, which
            is further transformed into a three-cluster state with a stationary cluster in the middle and two
            rotating clusters on the sides (see Figure~\ref{f.tri}\textbf{h}). An alternative scenario is shown in the
            last column of Figure~\ref{f.tri}. This time the AH bifurcation comes first and, therefore, we
            get a chimera state with two traveling coherent clusters on the sides. After the PF bifurcation, we arrive
            at the same three-cluster pattern as above (see Figure~\ref{f.tri}\textbf{i}).
            Finally, we break the even symmetry of $g$ by increasing $\sigma_3$. Without symmetry constraints,
            the point
            of triple intersection splits into three simple points $P_0, P_1,$ and $P_2$ (see Figure~\ref{f.tri+}\textbf{b}).
            Thus, we have a sequence of a PF bifurcation and two secondary bifurcations making clusters coherent and
            stationary one by one (see Figure~\ref{f.tri+}\textbf{c}). The eventual state is a stationary three-cluster
            state.

 \begin{figure*}
  \centering
  \begin{minipage}[b]{.98\textwidth}
  \textbf{a}\includegraphics[width=0.32\textwidth]{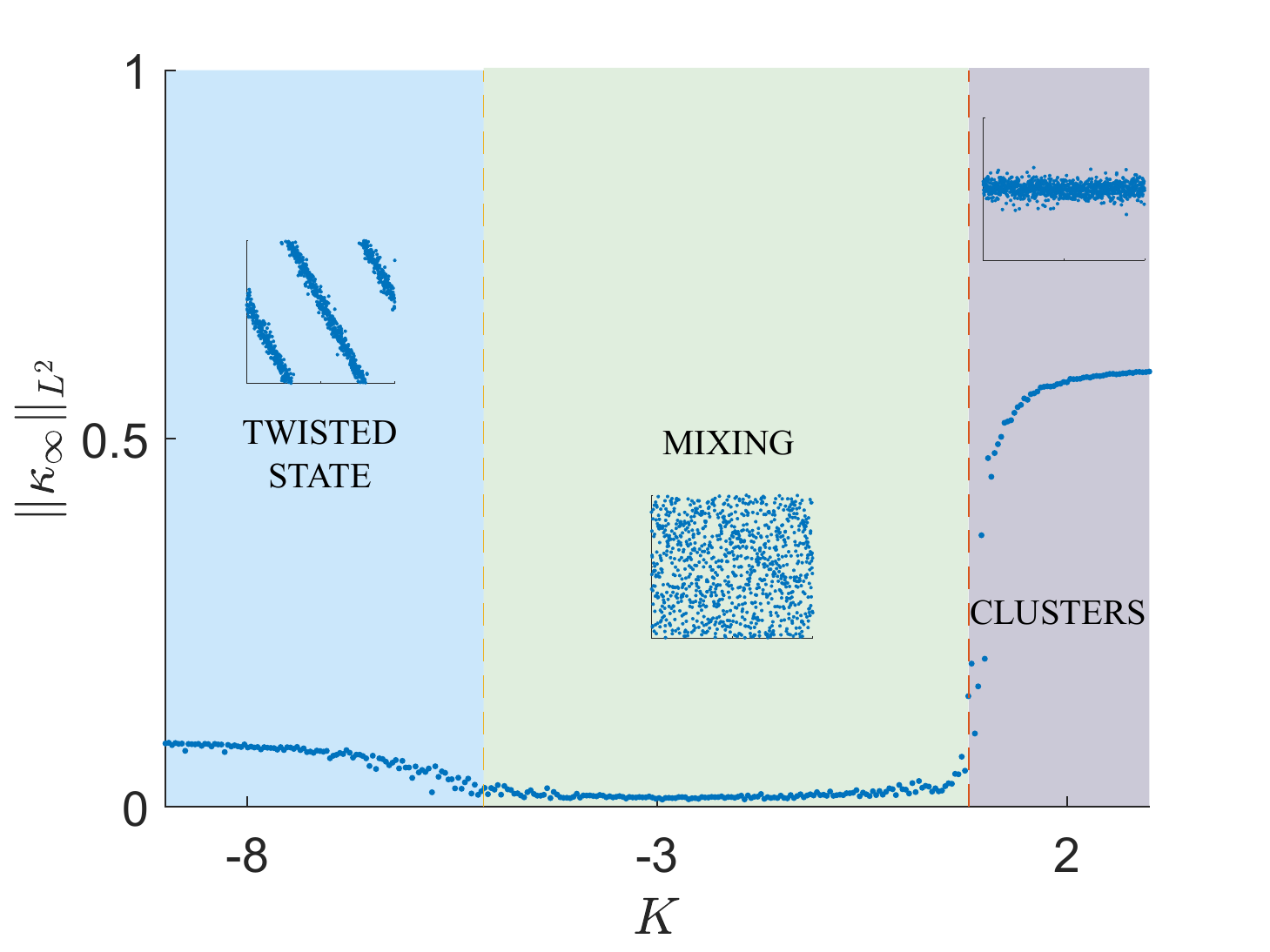}\hfill
  \textbf{b}\includegraphics[width=0.32\textwidth]{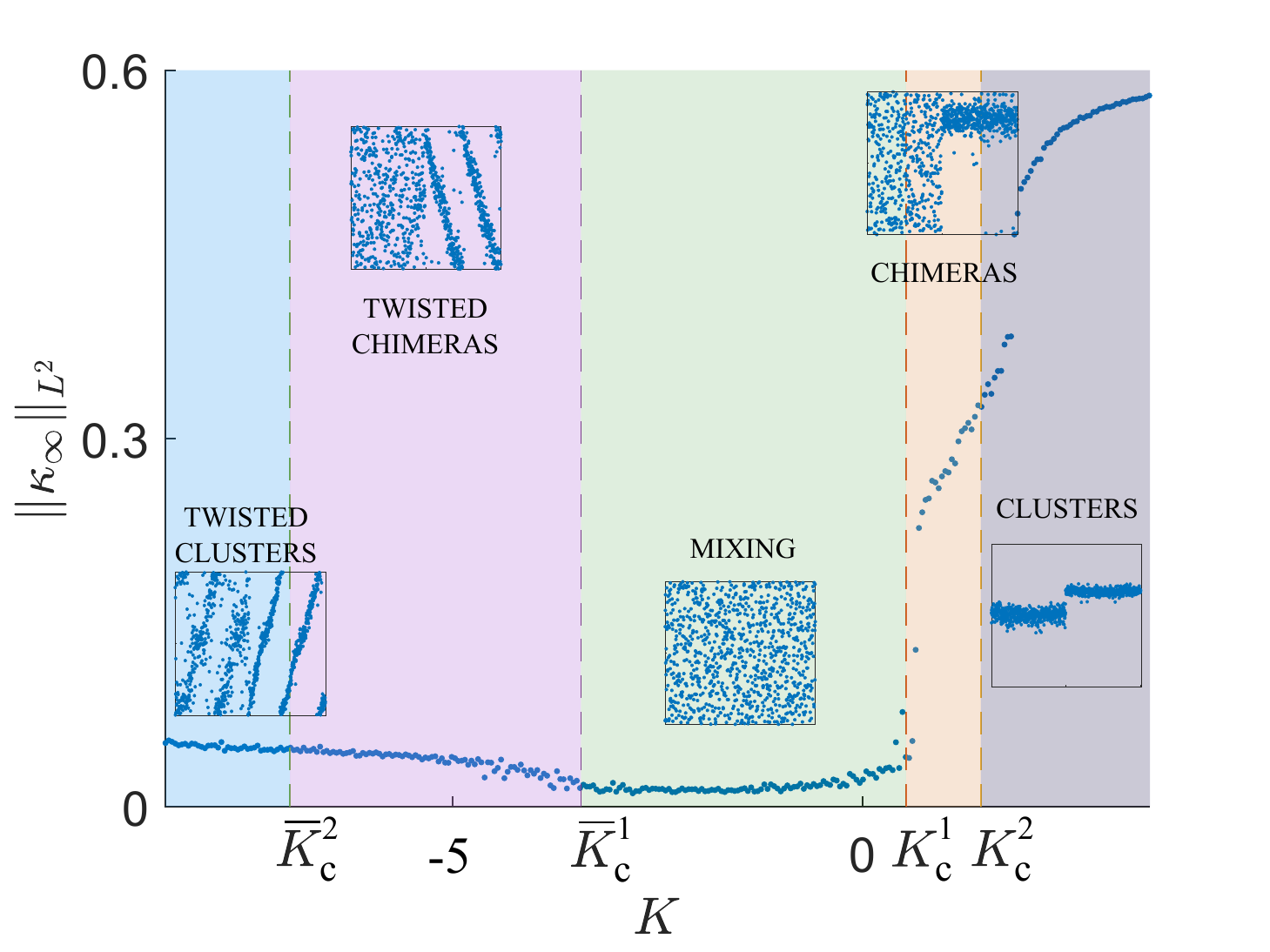}\hfill
  \textbf{c}\includegraphics[width=0.32\textwidth]{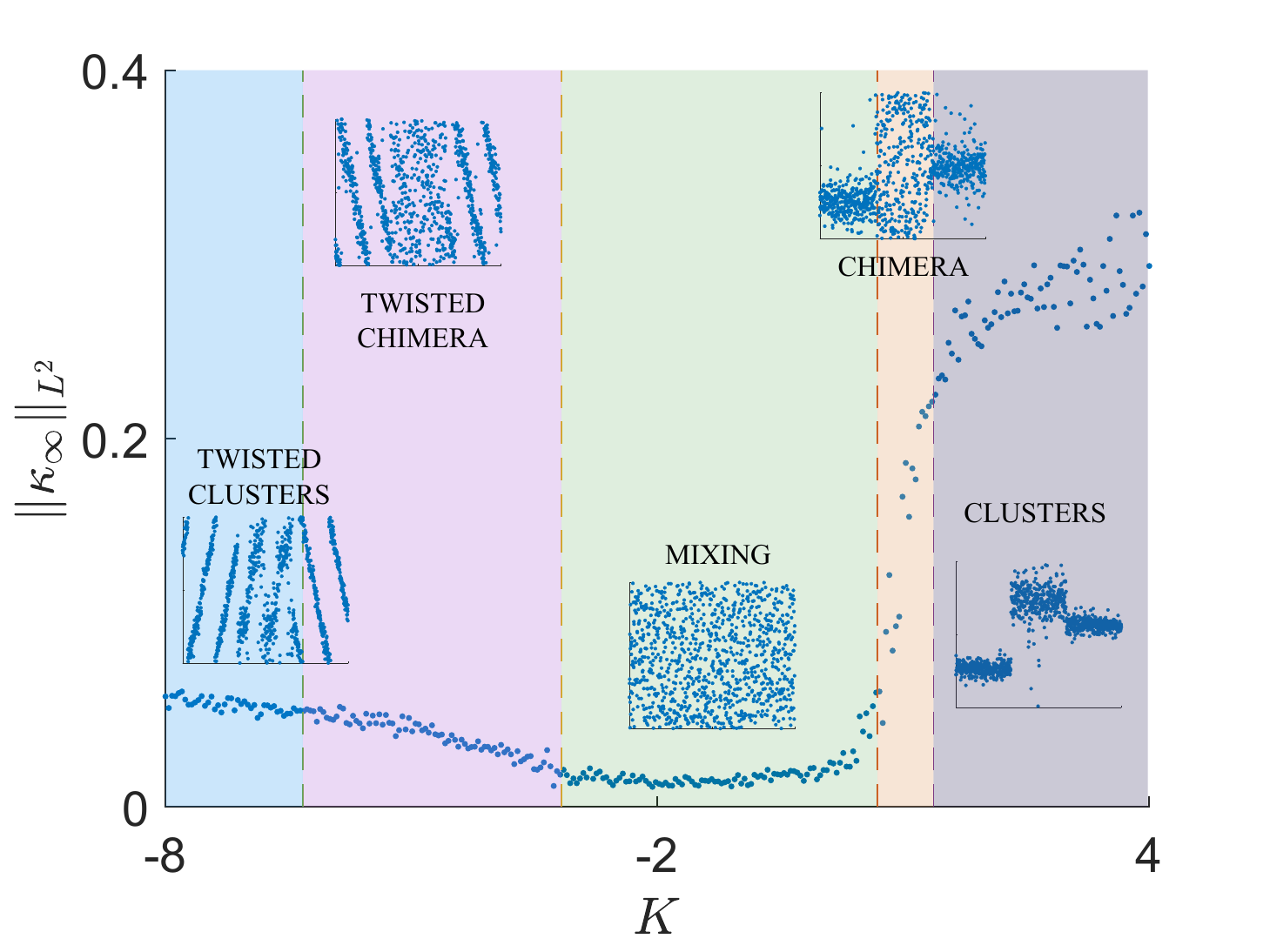}
  \caption{
    Bifurcation diagrams for the KM with nonlocal nearest-neighbor coupling and uni-, bi-, and tri-
    modal frequency distributions shown in plots \textbf{a}, \textbf{b}, and \textbf{c} respectively. In all simulations the nearest neighbor coupling range is $r=0.3$.
    In (\textbf{a}) the PF bifurcation leading to synchronization for positive $K$ has a counterpart
    for $K$ negative. The latter results in the emergence of stationary twisted states. The same
    principle applies to the bifurcation diagrams in \textbf{b} and \textbf{c}: bifurcations for positive
    $K$ have mirror images for negative $K$ with coherent structures superimposed on twisted states.
    For instance, diagram \textbf{b} features twisted chimera and clusters of traveling twisted
    states generated from the bimodal distribution in Fig.~\ref{f.bi}{\bf c}.
    Using the trimodal distribution from Fig.~\ref{f.tri}{\bf c}, we obtain the 3-clusters shown in \textbf{c}, two outer clusters contain traveling twisted states and the
    middle one presents a stationary twisted states. The cluster type (stationary vs traveling) is determined by the underlying bifurcation. The AH bifurcation results in a pair of traveling coherent structures, whereas PF bifurcation produces a stationary one (see text for details). Oscillators in the snapshots are reordered in order to easily distinguish the separate clusters.
}\label{f.W-bif}
\end{minipage}
\end{figure*}   

\section{Adding connectivity}\label{sec.connect}
Network connectivity can have a profound effect on the spatial organization
of chimera states.
In the previous section, we discussed bifurcation scenarios in the KM with all--to--all coupling,
i.e., for $W\equiv 1$. In this case, the largest eigenvalue of $\bW$ is $1$ and the corresponding
eigenfunction is $w\equiv 1$. There are no negative eigenvalues.  This has the following implications.
Mixing is stable for $K\in (-\infty, K_c)$ and patterns emerging at the bifurcation at $K_c$ are spatially
homogeneous, because $v_\lambda(\omega, x)$ in \eqref{v-lam} does not depend on $x$.
In general, $\bW$
may have  eigenvalues of both signs \cite{ChiMed19a}. In this case, along with the bifurcations
at positive
$K_c^1$ and $K_c^2$ identified above there are negative counterparts at $\bar K_c^2<\bar K_c^1<0$.
The eigenfunctions corresponding to negative eigenvalues of $\bW$ are no longer constant
and they endow the bifurcating patterns (clusters and chimeras) with a nontrivial spatial organization.
We refer an interested reader to \cite{CMM21, CMM18, ChiMed19a} for more details on the KM
with nonconstant $W$.

Suppose $\bW$ has eigenvalues of both signs
and denote the largest positive and smallest negative eigenvalues of $\bW$ by
$\mu^+$ and $\mu^-$ respectively. 
Then the region of stability of mixing is bounded $K\in (K_c^-, K_c^+),$ with $K_c^-<0<K_c^+$. Furthermore,
one of the eigenfunctions corresponding to $\mu^-$ or $\mu^+$ is not constant.
Thus, the patterns emerging from one
of the bifurcations will have spatial structure.

To illustrate these effects, we consider the KM with nonlocal nearest neighbor coupling. To this end,
let $W(x,y)=V(x-y)$, which is defined by
$$
V(x)=\1_{(-r,r)}(x), \quad \mbox{on} \; (-1/2, 1/2)
$$
and extended to $\R$ by periodicity. Here, $\1$ stands for the indicator function,
and $r\in (0,1/2)$ is a fixed parameter. Then
$$
\bW[f](x)=\int_0^1 V(x-y) f(y)dy.
$$
The eigenvalues of $\bW$ can be computed explicitly
$$
\mu_k=\int_0^1 V(x) e^{\pm 2\pi\iu kx}dx= \int_0^1 V(x) \cos\left(2\pi kx\right) dx, \quad k=0,1,2,\dots.
$$
The corresponding eigenfunctions are $w_k=e^{\pm 2\pi\iu kx}$.
The largest positive eigenvalue is $\mu^+=2r$ (cf.~\cite[Lemma~5.3]{ChiMed19a}).
By $k^\ast>0$ denote the value of $k$ corresponding to  the smallest negative eigenvalue
of $\bW$, $\mu_{k^\ast}$. The corresponding eigenfunctions are $e^{2\pi\iu k^\ast x}$ and
$e^{-2\pi\iu kx}$.

To explain the implications of the presence of the eigenvalues of both signs in the spectrum
of $\bW$, we first turn to the unimodal distribution. If $g$ is even and unimodal  then the region
of stability of mixing is a bounded interval $(K_c^-, K_c^+)$ with $K_c^-=\pi (g(0)\mu^-)^{-1}$
and $K_c^+=\pi (g(0)\mu^+)^{-1}$ \cite{ChiMed19a}. At $K_c^+$ we observe a familiar scenario
of transition to synchronization (Figure~\ref{f.W-bif}). At $K_c^-$ the situation is different.
The center subspace 
of the linearized problem in the Fourier space is spanned by
$$
v_{\mu^-}^{(1)} = \Upsilon_0(\omega) e^{2\pi\iu k^\ast x} \quad\mbox{and}\quad
v_{\mu^-}^{(2)} = \Upsilon_0(\omega) e^{-2\pi\iu k^\ast x}.
$$
In the solution space, we therefore expect that
$$
f(t,\theta,\omega,x)\sim
\operatorname{Re}\left( c_1+c_2 \Upsilon_0(\omega) e^{2\pi\iu (\pm k^\ast x-\theta)}\right), \quad
  c_1,c_2\in\C.
  $$
  For the PLS emerging at the bifurcation, we see that the structure encoded in $\Upsilon(\omega)$
  is now  superimposed onto a $k^\ast$--twisted state (Fig.~\ref{f.W-bif}\textbf{a}). The same principle
  applies to all other bifurcation scenarios, which we discussed for bi- and trimodal distributions in the
  previous section. Specifically, whenever a transition to coherence occurs whether in a cluster or in the
  entire population, the nascent coherent structure is superimposed onto a  twisted state.

  Plots \textbf{b} and \textbf{c} of Figure~\ref{f.W-bif} present bifurcation diagrams for families of
  bimodal and trimodal distributions. The bifurcations for positive $K$ analyzed in the previous sections 
  have  counterparts for negative $K$. The latter feature (traveling) twisted states every time the transition
  to coherence takes place. The velocity of the twisted states is determined by the corresponding eigenvalues of
  $\bT$ as before. The appearance of twisted states in this model is a consequence of anisotropic coupling.
  Whenever $W(x,y)=V(x-y)$ for some function $V$ on a unit circle, the eigenfunctions of
  $\bW$ are exponential functions $e^{2\pi\iu kx}$. By varying $W$, one can achieve a variety of
  spatial patterns born when mixing loses stability.

\section{Discussion}\label{sec.discuss}
\setcounter{equation}{0}

In this paper, we studied bifurcations in the KM with multimodal frequency distributions. We showed
that the loss of stability of mixing in this model leads to different patterns including stationary
and traveling clusters and chimera states. In structured networks these patterns acquire
additional spatial organization. In particular, bifurcations of mixing in the KM with nonlocal
nearest-neighbor coupling give rise to twisted chimera states with regions of coherent  behavior
organized as stationary or traveling twisted states. The combination of the linear stability analysis, Penrose
diagrams, and the spectral properties of the graph limit provide information 
about all essential features of the complex spatiotemporal patterns found in the KM after mixing loses statbility.
In particular, we were able to identify velocity distribution within chimera states as well the region
of their existence.

The type of the bifurcation determines the dynamical properties of the nascent patterns.
The PF bifurcation results in stationary clusters, whereas an AH bifurcation produces traveling clusters.
Furthermore, we described a codimension-2  and a codimension-3 PF-AH bifurcation, whose unfolding
contain transitions to stationary
and traveling clusters and chimera states in the KM with bi- and trimodal frequency distributions
(Figs.~\ref{f.bi-bif}, \ref{f.tri}, \ref{f.tri+}).  

To locate the bifurcations of mixing and subsequent (secondary) bifurcations leading to clusters or chimera
states we used Penrose's diagrams, which reduce the problem to the analysis of geometric and topological
properties of a closed critical curve. Once the bifurcations are found, the emerging patterns are
determined by the unstable modes, i.e., the eigenfunctions of the linearized operator corresponding to the
eigenvalues with zero real parts. In particular, the unstable modes determine the velocity distributions
within PLS and chimera states.

Our analysis of primary bifurcations of mixing relies on rigorous mathematical theory available for the
KM  \cite{ChiMed19a, ChiMed19b, Chiba2016Hopf, Die16}. Our treatment of the secondary bifurcations
should be considered as experimental. It predicts very well the emergence of chimeras, their statistical
properties, and the domain of existence, but the mathematical basis of these findings should be investigated
further. Nonetheless, our results show that the combination of the linear stability
analysis and the geometric method of Penrose provide a simple and effective way for understanding pattern
formation in the KM with multimodal frequency distribution. We believe that the bifurcation scenarios
described in this paper are relevant to other interacting particle systems. In particular, the same method
works well for the KM with inertia \cite{MM21}. It reveals similar and more sophisticated patterns for the model
with inertia. These results will be presented elsewhere.

Since their discovery chimeras have appealed to a broad community of mathematicians and physicists as
a stark example of highly heterogeneous structures produced by homogeneous systems.
An inquisitive reader might note that in contrast to the model in \cite{AbrStr06},
where all $\omega_i$'s are the same, in our case $\omega_i$'s are different, and so
the system is not homogeneous.
% One can go further
% and accuse us in using multimodal distributions promoting the formation of clusters and eventually
% chimeras.
In response, we note if  \eqref{KM} is rewritten as  
\begin{equation}\label{2KM}
  \begin{split}
  \dot\theta_i&=\omega_i +\frac{2K}{n} \sum_{j=1}^n a^n_{ij}\sin(\theta_j-\theta_i+\alpha),\\
  \dot\omega_i&=0,\quad i\in [n],
  \end{split}
\end{equation}
then the right-hand sides in all equations have the same form,
and heterogeneity enters only through the initial conditions. In this respect,
our setting is not different from that in \cite{AbrStr06}.

\vskip 0.2cm
\noindent

{\bf Acknowledgements.} This work was supported in part by NSF grant DMS 2009233 (to GSM).
Numerical simulations were completed using the high performance computing
cluster (ELSA) at the School of Science, The College of New Jersey. Funding of ELSA is provided
in part by NSF OAC-1828163. MSM was additionally supported by a
Support of Scholarly Activities Grant at The College of New Jersey.

\bibliographystyle{amsplain}
\def\cprime{$'$} \def\cprime{$'$}
\providecommand{\bysame}{\leavevmode\hbox to3em{\hrulefill}\thinspace}
\providecommand{\MR}{\relax\ifhmode\unskip\space\fi MR }
\providecommand{\MRhref}[2]{%
  \href{http://www.ams.org/mathscinet-getitem?mr=#1}{#2}
}
\providecommand{\href}[2]{#2}

\end{document}